\documentclass{emulateapj}
\usepackage{graphicx}
\usepackage{natbib}
\usepackage{color}

\usepackage{appendix}
\shortauthors{Nguyen et al.}
 
\begin{document}
 
\title{Extended Structure and Fate of the Nucleus in Henize 2-10}
 
\author{\mbox{Dieu D. Nguyen\altaffilmark{1}}}
\author{\mbox{Anil C. Seth\altaffilmark{1}}}
\author{\mbox{Amy E. Reines\altaffilmark{2}}}
\author{\mbox{Mark den Brok\altaffilmark{1}}}
\author{\mbox{David Sand\altaffilmark{3}}}
\author{\mbox{Brian McLeod\altaffilmark{4}}}

\altaffiltext{1}{Department of Physics and Astronomy, University of Utah, 115 South 1400 East, Salt Lake City, UT 84112, USA\\
{\tt dieu.nguyen@utah.edu}\\
{\tt aseth@astro.utah.edu}\\
{\tt denbrok@physics.utah.edu}}
\altaffiltext{2}{National Radio Astronomy Observatory, Charlottesville, VA 22903, USA, Einstein Fellow\\
{\tt areines@nrao.edu}}
\altaffiltext{3}{Department of Physics, Texas Tech University, 2500 Broadway St, Lubbock, TX 79409, USA\\
{\tt david.sand@ttu.edu}}
\altaffiltext{4}{Harvard-Smithsonian Center for Astrophysics, Harvard University, 60 Garden St, Cambridge, MA 02138, USA\\
{\tt bmcleod@cfa.harvard.edu}}
%%%%%%%%%%%%%%%%%%%%%%%%%%%%%%%%%%%%%%%%%%%%%%%%%%%%%%%%%%%%%%%%%%%%
% ABSTRACT SECTION
%%%%%%%%%%%%%%%%%%%%%%%%%%%%%%%%%%%%%%%%%%%%%%%%%%%%%%%%%%%%%%%%%%%%
\begin{abstract}
 We investigate the structure and nuclear region of the black hole (BH) hosting galaxy Henize~2-10. Surface brightness (SB) profiles are analyzed using Magellan/Megacam $g$- and $r$-band images.  Excluding the central starburst, we find a best-fit two component S\'ersic profile with $n_{\rm in} \sim 0.6$, $r_{\text{eff,in}} \sim$ 260~pc, and $n_{\rm out}\sim 1.8$, $r_{\text{eff, out}}\sim$ 1~kpc.  Integrating out to our outer most data point ($100\arcsec \sim 4.3$~kpc), we calculate $M_g=-19.2$ and $M_r=-19.8$.  The corresponding enclosed stellar mass is $M_{\star}\sim(10\pm3)\times10^9$~M$_\odot$, $\sim3\times$ larger than previous estimates.  Apart from the central $\lesssim$500~pc, with blue colors and an irregular morphology, the galaxy appears to be an early-type system.  The outer color is quite red, $(g-r)_0=0.75$, suggesting a dominant old population.  We study the nuclear region of the galaxy using archival Gemini/NIFS $K$-band adaptive optics spectroscopy and {\it Hubble Space Telescope} imaging.  We place an upper limit on the BH mass of $\sim10^7M_{\odot}$ from the NIFS data, consistent with that from the $M_{\rm BH}$-radio-X-ray fundamental plane.  No coronal lines are seen, but a Br$\gamma$ source is located at the position of the BH with a luminosity consistent with the X-ray emission.  The starburst at the center of Henize~2-10 has led to the formation of several super star clusters, which are within $\sim$100~pc of the BH.  We examine the fate of the nucleus by estimating the dynamical masses and dynamical friction timescales of the clusters.  The most massive clusters ($\sim 10^6~M_\odot$) have $\tau_{\rm dyn} \lesssim 200$ Myr, and thus Henize~2-10 may represent a rare snapshot of nuclear star cluster formation around a pre-existing massive BH.

\end{abstract}
%\slugcomment{To be accepted by the APJ}
\maketitle
%%%%%%%%%%%%%%%%%%%%%%%%%%%%%%%%%%%%%%%%%%%%%%%%%%%%%%%%%%%%%%%%%%%%
% INTRODUCTION SECTION
%%%%%%%%%%%%%%%%%%%%%%%%%%%%%%%%%%%%%%%%%%%%%%%%%%%%%%%%%%%%%%%%%%%%
\section{Introduction}\label{sec:intro}
The nuclei of galaxies are typically inhabitated by both massive black holes (BHs) and nuclear star clusters (NSCs). BHs have been found to be ubiquitous in galaxies more massive than the Milky Way, but at lower masses the picture is less clear.    BHs have been found in galaxies up to 100 times lower mass than the Milky Way \citep[][]{Verolme2002MNRAS, Barth2004, Greene2008ApJ, Seth2010ApJ, Reines2013ApJ, Reines2014ApJ}, but the occupation fraction of BHs in these galaxies remains uncertain \citep{Greene2012NatCo, Miller2014}.  NSCs are very common in fainter galaxies; about 75\% of galaxies with stellar masses between $5\times$10$^8$ and 10$^{11}$~M$_\odot$ have NSCs \citep{Boker2002AJ, Cote2006ApJ, Seth2008ApJ}.  The formation and evolution of BHs and NSCs is not well understood.  BHs are thought to originally form in the early universe from direct collapse or the remnants of Population III stars, and then grow through merger-induced or secular gas accretion, or through accretion of stellar material \citep{Volonteri2010A&AR}.  An indirect way of probing the formation of these objects is by examining how their mass scales with the properties of their host galaxies.  These scaling relationships have been studied for both BHs and NSCs, and their masses have been found to correlate strongly \citep{Ferrarese2006ApJ, Graham2012MNRAS, Leigh2012MNRAS}.  Initially, \citet[][]{Ferrarese2006ApJ} and  \citet[][]{Wehner2006ApJ} found that NSCs and BHs scale similarly with the mass of their host galaxies.  However, this claim has been disputed by more recent work \citep{Scott2013ApJ}.  

In the context of BH and NSC formation, Henize~2-10 is a particularly interesting object.  It is undergoing a major nuclear starburst with a star formation rate 1.9~$M_{\odot}\,\text{yr}^{-1}$ \citep{Mendez1999A&A, Engelbracht2005ApJ, Calzetti2007ApJ}.  A vast majority of this star formation is taking place in a very small region around the center of the galaxy with a diameter of 120~pc (3$\arcsec$).  There are many super star clusters (SSCs) $>10^5$~M$_\odot$ forming in this nuclear starburst, and these have been age-dated to be $<$5~Myr \citep{Chandar2003}.  A luminous nuclear X-ray point source was found by \citet[][]{Ott2005MNRAS.I, Ott2005MNRAS.II}.  \citet{Kobulnicky2010ApJ} suggested it may be an intermediate-mass BH or ultraluminous X-ray source.  Stronger evidence for the existence of an accreting massive BH was recently found by \citet[][]{Reines2011}, who showed that the X-ray source was coincident with a radio source.  The luminosities of this source are consistent with a $\sim10^6M_{\odot}$ BH lying on the fundamental plane \citep{Merloni2003MNRAS}.  \citet[][]{Reines2011} find that alternative explanations for this source are not plausible, and conclude that there is likely an accreting BH. The source is not coincident with any of the known SSCs and parsec-scale non-thermal radio emission is detected from the source with very long baseline interferometry observations \citep{Reines2012ApJ}.

In this paper, we focus on three aspects of what this interesting galaxy can tell us about BH and NSC formation.  First, we examine the extended structure and stellar mass of the galaxy and place it in a broader context.  Next, we examine the kinematics of the nuclear region, and use this information to look at whether an NSC can form from the existing population of SSCs found in the nucleus.  Finally, we examine the infrared spectra at the location of the putative BH for any signs of BH accretion visible in the NIR. 

Henize~2-10 is a blue compact dwarf or Wolf-Rayet galaxy, with an absolute $B$-band magnitude of -19.07 ($6.5\times10^9 L_\odot$) \citep{Micheva2013A&A}.  Due to its low Galactic latitude ($b=8.6^\circ$) and significant foreground contamination, He~2-10's larger scale structure has global structural properties are not fully understood.  Deep imaging of the galaxy in the NIR by \citet{Noeske2003A&A} and the optical by \citet{Micheva2013A&A} suggest the galaxy has an extended red halo around the central starburst.  \citet{Corbin1993AJ} published a de Vaucouleurs SB profile fit to V band data from 6.75$\arcsec$ to 40$\arcsec$; here we present fits to higher quality imaging data covering a larger radial extent.  The total dynamical mass of the galaxy was estimated from poorly resolved HI kinematics to be 2.9$\times 10^9/{\rm sin}^2\,i\;M_\odot$ within 2.1~kpc and the total gas mass is $\sim 7 \times 10^8 M_\odot$ \citep{Kobulnicky1995AJ,Kobulnicky2010ApJ}. Here we investigate the extended structure of Henize~2-10 by fitting S\'ersic functions to its SB profiles based on ground-based imaging data in $g$- and $r$-bands.  

Many kinematic observations have also been made of the starburst region of Henize~2-10.  The gas and stellar kinematics show dramatically different components.  The gas is clearly rotating within the central 8$\arcsec$ as traced by HI, CO and NIR molecular lines \citep[][]{Kobulnicky1995AJ, Santangelo2009A&A, Cresci2010A&A}.  The rotation speeds are as high as 30 - 80~km~$\text{s}^{-1}$, while the velocity dispersion of this component is typically $<$ 50 km $\text{s}^{-1}$.  On the other hand, the stellar component within the central 6$\arcsec$ is non-rotating and dispersion dominated \citep[][]{Marquart2007A&A}.  The velocity dispersion of this component is $\sim$ 45 km $\text{s}^{-1}$ which implies a dynamical mass of $\sim6\times10^8$~M$_\odot$ within the central 6$\arcsec$.  We investigate the global and internal kinematics of the SSCs within the nucleus using Gemini/NIFS observations.  

Because of the coexistence of a BH and SSCs in the central 100~pc of Henize 2-10, we expect they may migrate to the center due to dynamical friction and form a NSC.  This idea was first introduced to explain the nucleus of M31 \citep{Tremaine1975}, and more recently has been re-examined in the context of dwarf elliptical galaxies \citep{Lotz2001ApJ}, spiral galaxies \citep{Milosavljevic2004ApJ} and in the presence of a massive BH \citep{Antonini2012ApJ, Antonini2013, Antonini2014}. In order to examine this phenomenon in the context of Henize~2-10, we examine the SSC masses, combining the NIFS and archival {\it Hubble Space Telescope (HST)} imaging to model their dynamical friction timescales. 

This paper is organized into 6 sections. In Section 2, we describe the observations and data reduction.  The extended structure of Henize~2-10 will be discussed in Section 3.  Our results on the future of the SSCs at the center of Henize~2-10 are presented in Section 4. In Section 5, we examine the NIR spectroscopy data and report a possible signature of the BH accretion based on Br$\gamma$ luminosity. We conclude in Section 6.  We assume a distance to Henize~2-10 of 9 Mpc \citep{Mendez1999A&A}; the physical scale assuming this distance is $\sim$ 43~pc~$\arcsec^{-1}$.  Unless otherwise indicated, all quantities quoted in this paper have been corrected for a foreground extinction $A_V = 0.306$  \citep[$A_g = 0.369$, $A_r = 0.255$;][]{Schlafly2011ApJ}.
%%%%%%%%%%%%%%%%%%%%%%%%%%%%%%%%%%%%%%%%%%%%%%%%%%%%%%%%%%%%%%%%%%%% 
% DATA SECTION 
%%%%%%%%%%%%%%%%%%%%%%%%%%%%%%%%%%%%%%%%%%%%%%%%%%%%%%%%%%%%%%%%%%%% 
\section{Data and Data Reduction}\label{sec:data}
%%%%%%%%%%%%%%%%%%%%%%%%%%%%%%%%%%%%%%%%%%%%%%%%%%%%%%%%%%%%%%%%%%%%
\subsection{Magellan/Megacam data}
Images of Henize~2-10 were taken on Nov. 21, 2011 with Megacam on the Magellan/Clay telescope. The seeing was poor (1.5-2$\arcsec$), but the night was photometric. Five 100s exposures were taken in $g$-, $r$-, and $i$-band at an airmass of $\sim$~1.15; the i band data suffered from fringing and poor sky subtraction and thus was not used in our analysis. The images were dithered to remove the chip gaps. Final, flat-fielded images were created by the pipeline at the Smithsonian Astrophysical Observatory Telescope Data Center; we made further corrections to the sky subtraction on each chip to ensure a flat field. Specifically, we fit a gradient to the residual background in each chip to a linear function along the horizontal direction and subtracted this off. We then matched the background in neighboring fields to ensure as flat a field as possible before combining the images using {\tt Swarp} \citep{Bertin2002ASPC}.  Flux calibration was obtained through observations of equatorial SDSS fields in each band, and are thus on the SDSS system.
%%%%%%%%%%%%%%%%%%%%%%%%%%%%%%%%%%%%%%%%%%%%%%%%%%%%%%%%%%%%%%%%%%%%
\subsection{Gemini/NIFS data} 
Gemini/NIFS data of Henize~2-10 was taken in April 2010 using Altair adaptive optics for a program (PI: Usuda) to examine the SSCs and gas conditions within the galaxy.  We downloaded the K-band data and accompanying calibration observations from the Gemini Science Archive.  The total data included 9 on-source and 9 off-source exposures, each with an exposure time of 300s.  The data were taken at airmass between 1.5 and 1.9.  Data reduction was performed as outlined by \citet[][]{Seth2010ApJ}, with telluric calibration performed using an A0V star (HIP45037) taken at similar airmass.  Flux calibration of the spectroscopic cube was performed by using a zeropoint derived from telluric stars and a 2MASS $K_s$ filter response curve; the flux calibration is expected to be good to $\sim$~10\%.  Final cubes were constructed from 8 of the 9 sky-subtracted on-source cubes with good image quality.  The line spread function in each pixel was determined by using a sky cube created with the same dither pattern as the science data cube; the median FWHM is 4.15\AA (R $\sim$ 5500), with a variation from 3.4 and 4.9\AA~across the map.   The FWHM of the AO corrected PSF in the final cube is quite good; the brightest SSC, which we refer to as CLTC1 following \citet{Chandar2003}, has a measured FWHM of $\sim0\farcs24$.  From {\it HST/HRC} images (\S2.3), we measured a true FWHM of $\sim0\farcs19$, suggesting the true PSF is $\sim0\farcs15$ FWHM (3 pixels in our final cubes).   The signal-to-noise ratio (SNR) per pixel ranges from 5-65 across the cube.  The NIFS data cubes were astrometrically aligned to previous measurements using {\it HST} data aligned to 2MASS astrometry as in \citet{Reines2011}.  The absolute error in astrometry is thus $\lesssim0\farcs1$, and the astrometric error relative to the nuclear X-ray source is 0$\farcs$13 in RA and 0$\farcs$33 in Dec, while the alignment of the nuclear radio source is $<0\farcs$1 \citep[supplementary materials;][]{Reines2011}.  All velocities are corrected to the barycentric velocity; the correction applied was -20.35~km~$\text{s}^{-1}$.
%%%%%%%%%%%%%%%%%%%%%%%%%%%%%%%%%%%%%%%%%%%%%%%%%%%%%%%%%%%%%%%%%%
\subsection{HST data}

\begin{figure}[h]
     \centering
      \epsscale{1.25}
          \plotone{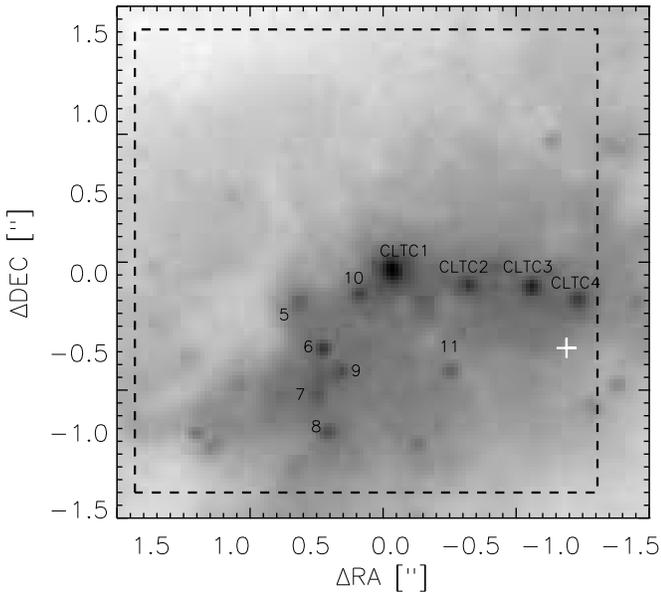}
     \caption[SCP06C1]{ {\it HST ACS/HRC} $F814W$ of the central region of Henize 2-10. The 11 SSCs in our sample are labeled. CLTC1-4 have been previously identified in \citet[][]{Chandar2003}; the other clusters are identified in this paper and are labeled with numbers 5-11 (see Table~3 for RA and Dec positions).  The dashed black box is the Gemini/NIFS field of view. The white cross indicates the BH's position. The position is given relative to RA = $08^h36^m15^s.199$ , Dec = $-26^{\circ}24^{\prime}33^{\prime\prime}.62$. }
\label{SSCs}
\end{figure}

We use {\it HST} data in two filters in this paper: (1) {\it ACS/HRC} data in the {\it F814W} filters taken in November 2005 (Program ID: 10609, PI: Vacca) and (2) {\it NICMOS NIC2} data in the {\it F205W} filter taken in October 2006 (Program ID: 10894, PI: Johnson).  Due to the high resolution of the {\it F814W} data, we used it to select the brightest star clusters in the starburst region as shown in Fig.~\ref{SSCs}.  These images are used in combination with the Gemini/NIFS data to determine the masses of SSCs and their possible dynamical friction timescales  of forming an NSC at the center of Henize~2-10 (\S4). 
%%%%%%%%%%%%%%%%%%%%%%%%%%%%%%%%%%%%%%%%%%%%%%%%%%%%%%%%%%%%%%%%%%
\subsection{Stellar Kinematics of Gemini/NIFS data} 
We used the procedure described in \citet[][]{Seth2010ApJ} to derive stellar kinematics in the central starburst region of Henize~2-10. Before deriving kinematics, we bin spectra using the Voronoi binning method \citep{Cappellari2003} to ensure a signal-to-noise, SNR, of $\sim$25 in each bin. Next, we determined the radial velocity, $v$, the velocity dispersion, $\sigma$, the skewness, $h_3$, and the kurtosis, $h_4$, in the wavelength region from 2.28~$\mu$m to 2.40~$\mu$m including the position of CO absorption bandheads by using the PPXF code of \citet[][]{Cappellari2004PASP}. This code requires stellar templates; we used high resolution templates from \citet[][]{Wallace1996ApJS} of eight stars with spectral types between G and M and including all luminosity classes. The program finds the best matching template from the set of templates, convolves them with  the line spread function determined from fits to sky lines in each pixel, and finds the best light-of-sight velocity distribution (LOSVD) which is described in term of Gauss-Hermite series.  Errors on the LOSVD are calculated from Monte Carlo simulations adding in Gaussian random error to each pixel and redetermining the velocities.  These radial velocity and dispersion errors range from 0.3~km~$\text{s}^{-1}$ to 25~km~$\text{s}^{-1}$; we note that systematic errors due to sky and background subtraction, telluric correction, and template mismatch are likely of order a few km~$\text{s}^{-1}$; more discussion of the systematic errors is presented in \S4.2. Based on visual inspection, errors in dispersion larger than 20\% were unreliable, and therefore we eliminate these bins from our analysis.  Although we do not cover the full central region, the median velocity in the higher SNR regions of our maps is 876~km~$\text{s}^{-1}$ or $\sim$4~km~$\text{s}^{-1}$ offset from the systemic velocity of 872$\pm$6~km~$\text{s}^{-1}$ given by \citet[][]{Marquart2007A&A}.  We therefore use the Marquart systemic velocity value throughout this paper.  

\begin{figure}[h]     
      \centering
          \epsscale{2.4}
          \plottwo{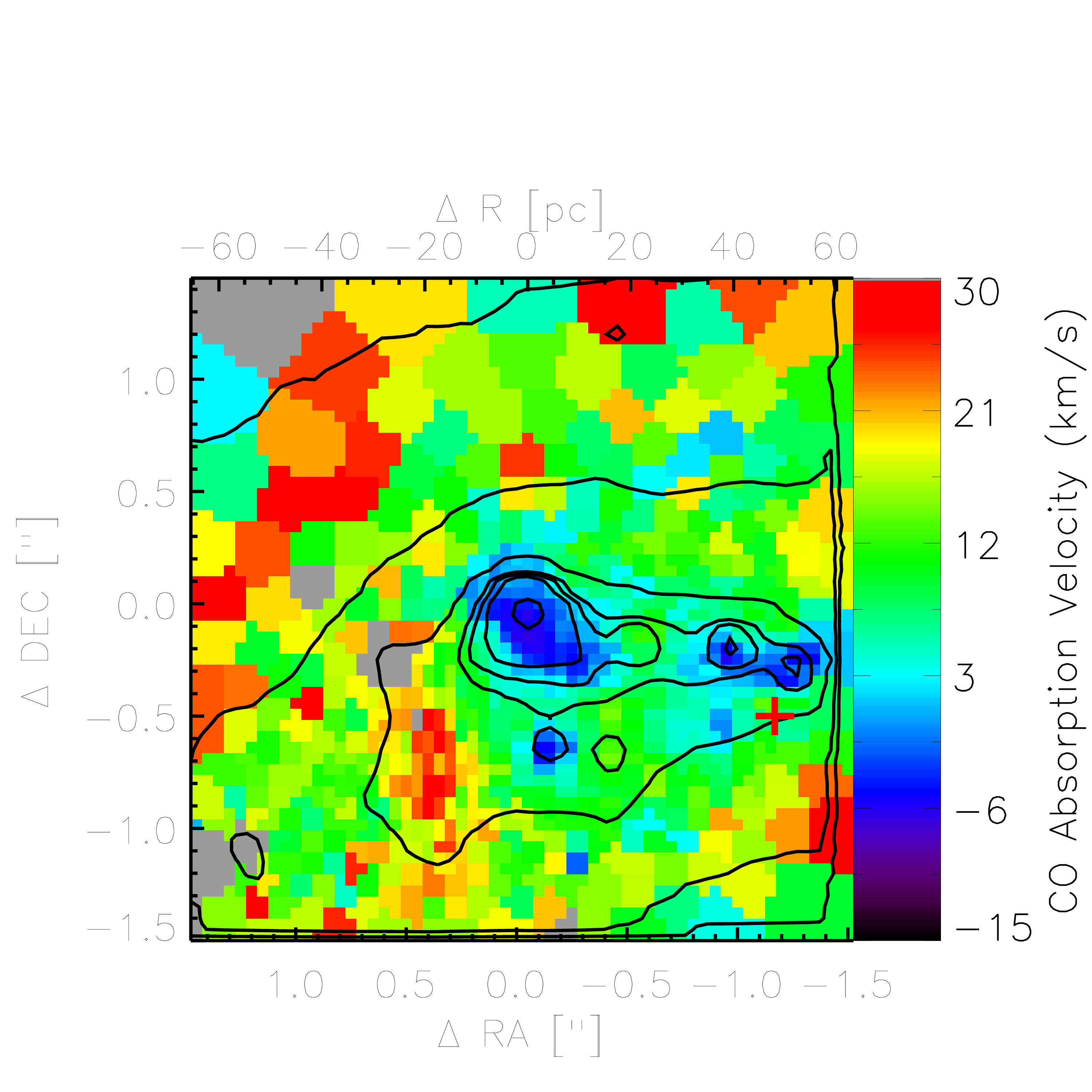}{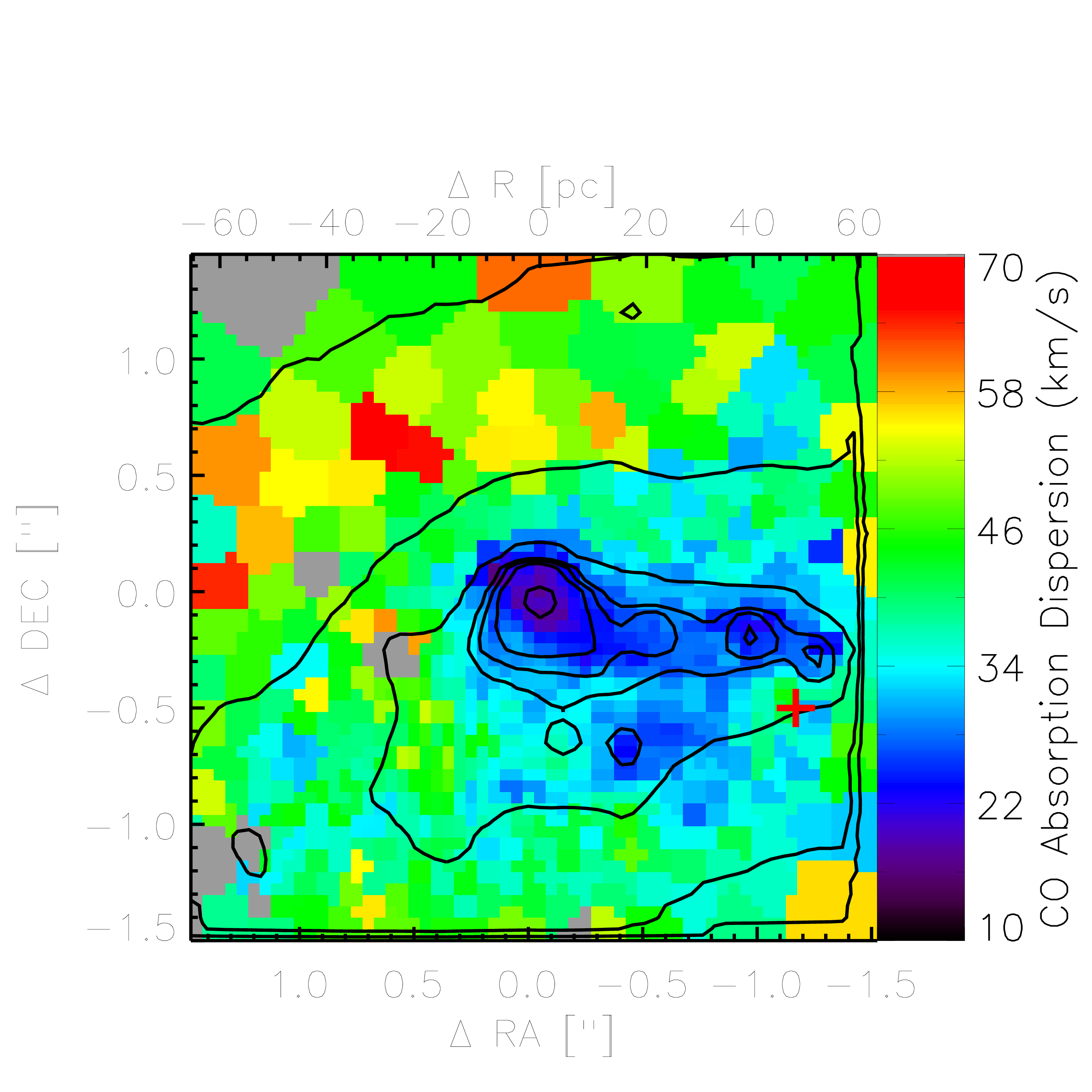}  
     \caption[SCP06C1]{ Velocity map (\textit{Upper Panel}) and dispersion map (\textit{Lower Panel}) of the stellar component derived the from CO absorption line bandhead.  The radial velocity is shown relative to the systemic velocity derived by \citet[][]{Marquart2007A&A} of 872$\pm$6 km~$\text{s}^{-1}$.  The contours show the K band image continuum with contours at surface brightness of 15.7, 15.9, 16.1, 16.6, 17.7 mag/$\arcsec^2$. The red cross marks the location of the BH. The positions are given relative to CLTC~1 (Table 3).}
\label{COfig}
\end{figure}

Fig.~\ref{COfig} shows the velocity and dispersion maps of the stellar component derived from the CO bandhead.  We note that our data do not cover the entire nuclear region of Henize~2-10; the expected location of the BH is near the western edge of our maps and is marked. by a red cross in Fig.~\ref{COfig}.  No peak in total intensity or change in dispersion or velocity is seen at the location of X-ray and radio source associated with the BH. 

\begin{figure}[h]     
      \centering
          \epsscale{1.3}
          \plotone{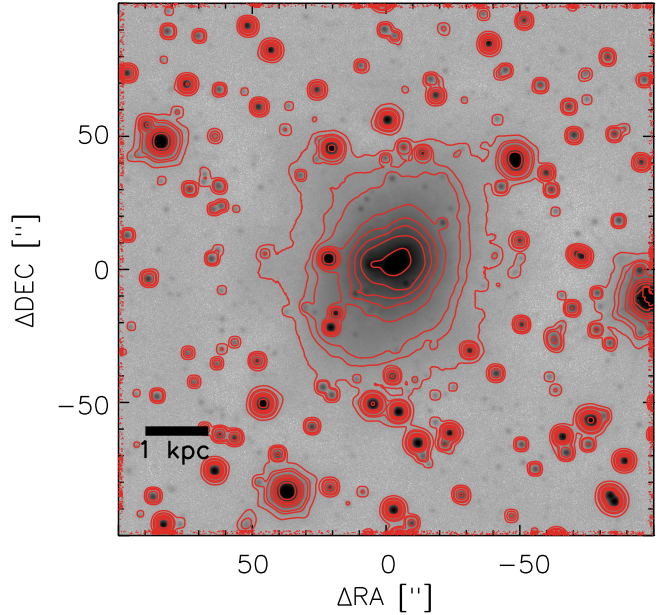} 
     \caption[SCP06C1]{ Contour plot based on the Megacam $r$-band image of Heize~2-10. The red contours are at a surface brightness $\mu_r=$18.5, 19.5, 20.0, 20.7, 21.5, 22.5, and 23.5 mag/$\arcsec^2$.}
\label{contourfig}
\end{figure}

The kinematic maps show two distinct regions. The stars clusters are blue shifted relative to the surrounding stars by 20~km~$\text{s}^{-1}$ to 30~km~$\text{s}^{-1}$. The dispersions in these regions are low, $\lesssim 30$~km~$\text{s}^{-1}$.  The clusters are apparently rotating in the same sense as the gas component \citep{Cresci2010A&A}, with an amplitude somewhat less than the gas.  Outside of the clusters, the kinematics have a higher dispersion of $\sim$~45~km~$\text{s}^{-1}$ and don't appear to have strong rotation.  The kinematics of these non-cluster stars are consistent with the lack of rotation and $\sim$~45~km~$\text{s}^{-1}$ dispersion observed over the central 6$\arcsec$ region by \citet[][]{Marquart2007A&A}. Thus it appears the dispersion profile of the non-cluster component within the inner $\sim$6$\arcsec$ is nearly flat.   We note that in the central region where we are observing, the seeing limited measurements of \citet[][]{Marquart2007A&A} find a dispersion of $\sim$28~km~$\text{s}^{-1}$, likely due to the influence of the lower dispersion star clusters.  We discuss dispersion measurements of the most massive cluster, CLTC1, in \S4.2.  

\begin{figure}[h]
      \centering
          \epsscale{1.23}
          \plotone{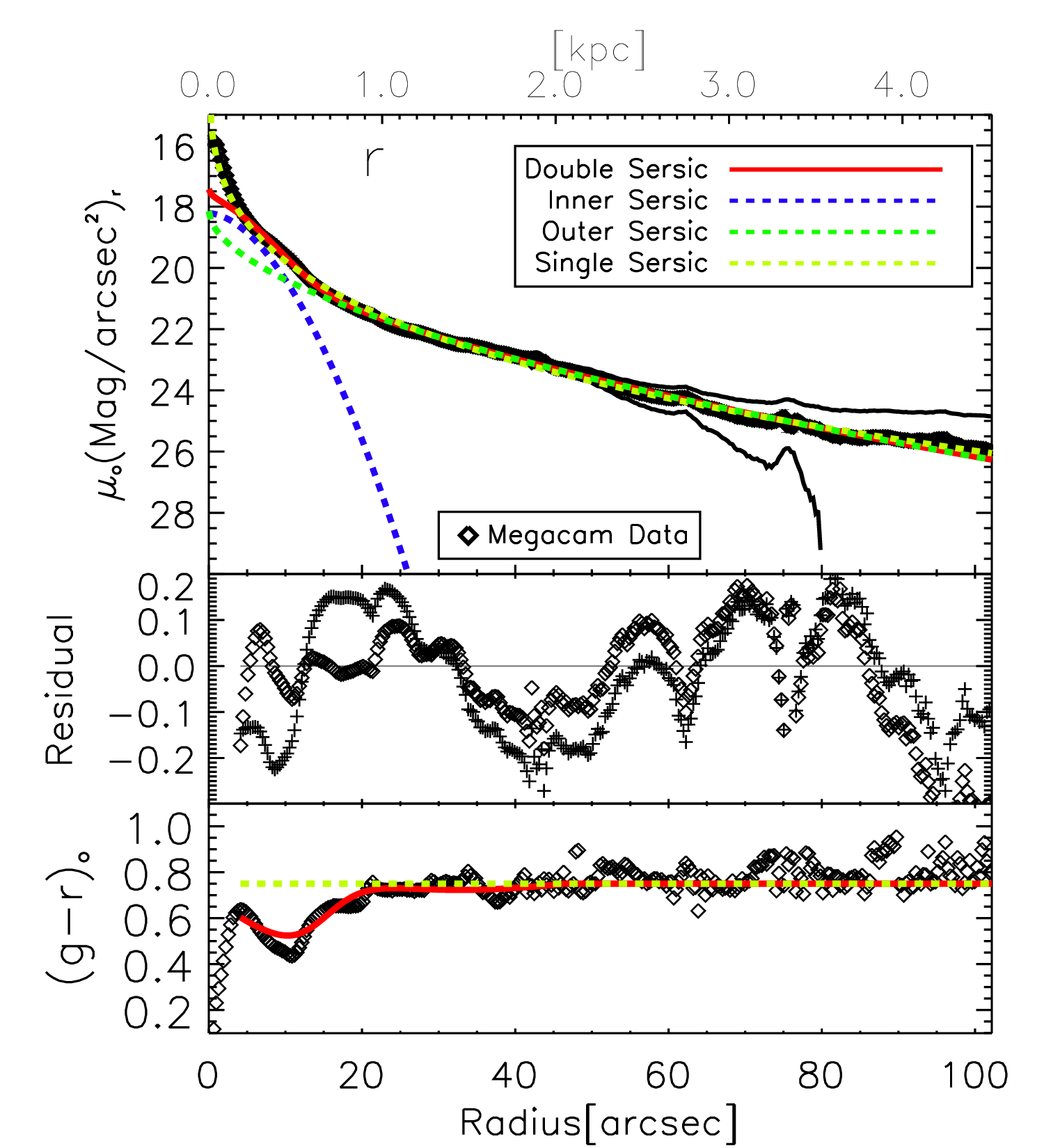}
     \caption[SCP06C1]{\textit{Upper Panel:} The $r$-band SB profile of Henize~2-10. Data points show the measured profile with the black lines showing the systematic uncertainty due to sky subtraction.  The best-fit double-S\'ersic model is shown in red and the best-fit single-S\'ersic model is shown as a dashed yellow line. The individual components of the Double-S\'ersic model are shown as dashed blue and green lines. We note that the inner 2$\arcsec$ region is excluded from the fit and the profiles in this region are extrapolated from the best fits. \textit{Middle Panel:} The SB profile residuals after subtraction of the double-S\'ersic (diamonds) and the single-S\'ersic (pluses) model profiles. \textit{Bottom Panel:} The color profile of Henize~2-10.  The black diamonds are the data points from the Megacam data, the red solid line indicates the double-S\'ersic component fit, and the yellow dashed line indicates the single-S\'ersic component fit applied to both $g$- and $r$-bands.}
\label{sersicfig}
\end{figure}

\begin{figure}[h]
     \centering
      \epsscale{1.24}
          \plotone{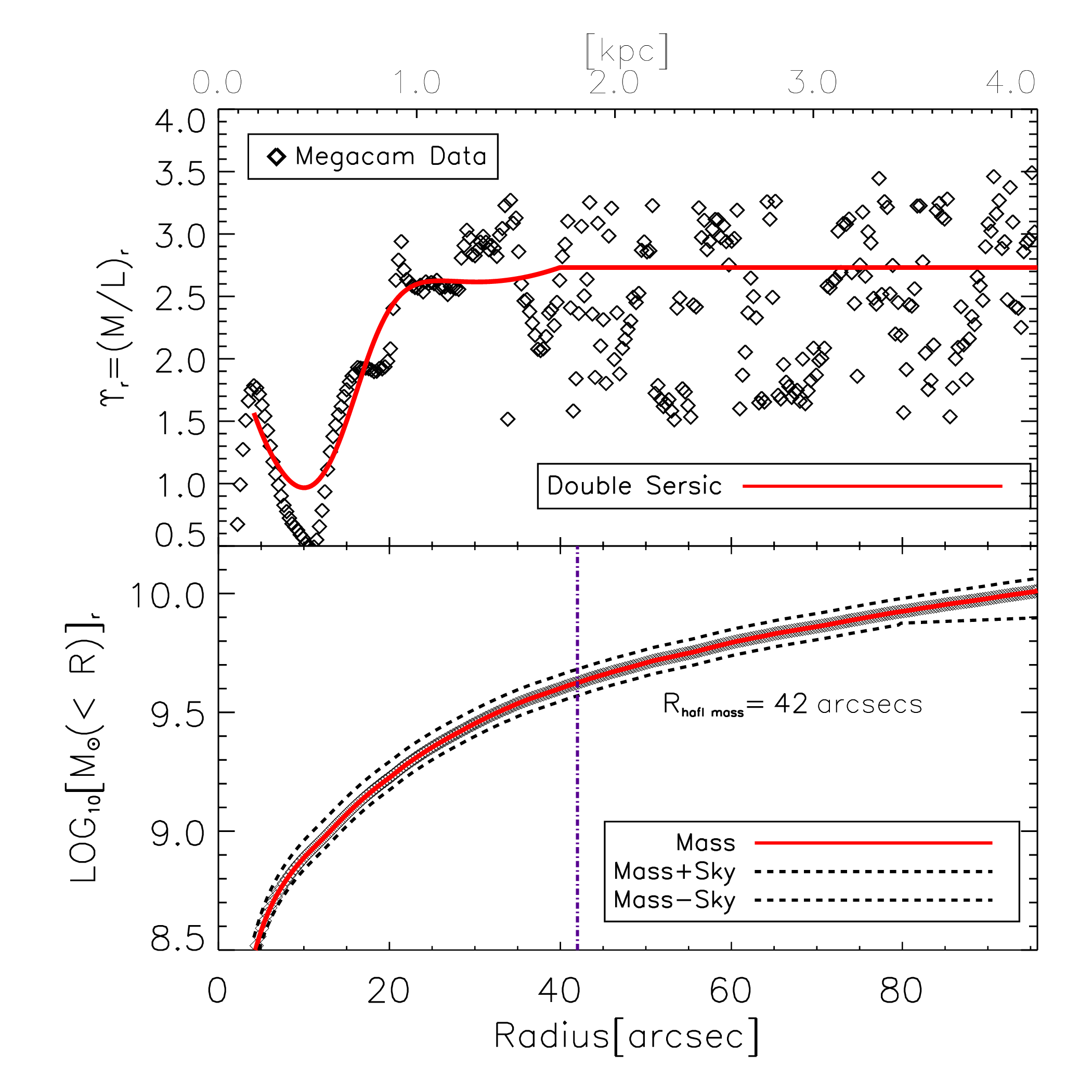}
     \caption[SCP06C1]{\textit{Upper Panel:} The mass-to-light profile in the $r$-band determined from the colors using the relation of \citet[][]{Zibetti2009MNRASZ}, the black diamonds indicates the measured colors while the red solid line shows the results from the two component S\'ersic model profile. \textit{Bottom Panel:} The cumulative mass within radius R as determined from the $r$-band image.  The purple vertical line shows the half mass radius. The black diamonds are the mass profile from the data and the red solid line is the predicted mass profile from two component S\'ersic fit. The black dashed lines show the uncertainty in the mass due to sky subtraction.}
\label{m2l_massfig}
\end{figure}
%%%%%%%%%%%%%%%%%%%%%%%%%%%%%%%%%%%%%%%%%%%%%%%%%%%%%%%%%%%%%%%%%%%%%%%%%%%%%
% LARGE SCALE STRUCTURE OF HENIZE 2-10 SECTION
%%%%%%%%%%%%%%%%%%%%%%%%%%%%%%%%%%%%%%%%%%%%%%%%%%%%%%%%%%%%%%%%%%%%%%%%%%%%%
\section{Extended Structure}\label{sec:structure}
%%%%%%%%%%%%%%%%%%%%%%%%%%%%%%%%%%%%%%%%%%%%%%%%%%%%%%%%%%%%%%%%%%
\subsection{SB Profiles}
To examine the extended structure and nature of Henize~2-10, we constructed SB profiles from the $g$- and $r$-band Magellan/Megacam images (\S2.1).  We first masked out all foreground stars and background galaxies by using $\tt{SEXTRACTOR}$ \citep{Bertin1996A&AS} to determine the shape, size, and position of each source.  Each object's size was increased by a factor of 2 while masking.  The SB profiles were then derived using the $\tt{IRAF\;ellipse}$ routine to extract fluxes, major semiaxis, etc. While extracting the fluxes, we use the best fit position angle and ellipticity from \citet[][]{Micheva2013A&A}; the ellipticity is fixed at $e=0.15$, and the position angle is fixed at $PA=28^{\circ}$.  The Magellan $r$-band images with the contours are presented in Fig.~\ref{contourfig}.  The $r$-band SB profile is shown in Fig.~\ref{sersicfig}. The total luminosity of the galaxy in $g$- and $r$-band between 0$\arcsec$ and 100$\arcsec$ is $\sim5.8\times10^{9} L_{\odot}$ and $\sim6.3\times10^{9} L_{\odot}$. The effective (half-light) radius of the overall galaxy is 30$\farcs$0 (1.3~kpc) and 32$\farcs$0 (1.4~kpc) in $g$- and $r$- band, respectively.

The central most region of Henize~2-10 is dominated by a young starburst; to focus on the extended structure and minimize the affect of this starburst we only fit the profile beyond 2$\arcsec$.  We note that because our PSF FWHM is smaller than this radius, the affects of the PSF on our SB profile fits are minimal.  The outer boundary of our fits are set by the error in our sky subtraction; this uncertainty in the sky subtraction is shown by black solid lines in Fig.~\ref{sersicfig}; beyond 70$\arcsec$ the sky subtraction makes the SB profile less reliable.  The outer parts were fit by a double-S\'ersic profile for both $g$- and $r$-images from $2\arcsec$ (86~pc) to $100\arcsec$ (4.3~kpc) from the center. Here we used the non-linear least squares {\tt IDL MPFIT} function to implement the fitting procedure. Henize 2-10's (SB) profile in both bands is best fit by two S\'ersic components, an inner component with $n \sim 0.6$ and $r_{\text{eff}} \sim 6\arcsec$ (260~pc) and an outer component with $n \sim 1.8$ and $r_{\text{eff}} \sim 25\arcsec$ (1~kpc).  The full results and errors in each band are shown in Table~\ref{table:MorphoHe2-10}.  Fig.~\ref{sersicfig} shows both the best double- and single-S\'ersic fit profiles in r-band. The single-S\'ersic fit to the r-band image provides a worse fit, but has $n = 4$ (a de Vaucouleurs profile) and $r_{\text{eff}} = 12.5\arcsec$ (538 pc).  In addition to having larger residuals, the single-Sersic fit (1) does not well match the observed color profile, (2)  the $\chi^2$ are $\sim$25\% larger than that of the double-S\'ersic profiles, and (3) over-predicts the amount of light even with the overlying starburst component at the center of the galaxy.  Therefore we prefer the double-Sersic fit and use this fit throughout the paper. We note that while \citet{Corbin1993AJ} do not provide the effective radius of their fits, both our single and double-Sersic fits are better fits to the data than the de Vaucouleurs profile fit they show in their Fig. 3 of \citet{Corbin1993AJ}.  Specifically, their fit is poor at radii less than 14$\arcsec$.  

 To estimate the effect of our choice of inner and outer boundaries on the fit, we varied the inner boundary from 2$\arcsec$ to 10$\arcsec$ and outer boundary from 70$\arcsec$ to 110$\arcsec$ for both $g$- and $r$-images; this resulted in a variation of the inner S\'ersic index and effective radius of 1\% and the outer fitted S\'ersic SB profile less than 5\%. Our results therefore appear robust to the range over which the profile is fitted.   The profiles are fit well by the two S\'ersic components model, with residuals less than $\sim$ 20\% except for an outer starburst component seen at a radius of $\sim 10\arcsec$.  The colors throughout the region (except for the starburst) are consistent with a $>$1 Gyr old population and thus the SB profile we fit appears to describe the older underlying component of Henize~2-10; this is especially true for the outer component, which dominates beyond $\sim$10$\arcsec$.  

To test our results, we compared our profiles with those in Bessel-Johnson $UBVHK$ filters \citep{Micheva2013A&A} using conversions between Sloan and Johnson filter sets \citep{Fukugita1996AJ,  Smith2002AJ, Jester2005AJ}.  The transformed profiles agree from the center to 90$\arcsec$ to within 10\%, beyond this distance, a somewhat larger systemic discrepancy of $\sim$20\%  exists between the $g$- and $B$-band.  This discrepancy does not affect our profile fits which are performed only out to 100$\arcsec$.  We also fit all 5 filters $UBVHK$ in the same range of radii from 2$\arcsec$ to 100$\arcsec$ with the similar double-S\'ersic profile which we obtained from our $g$- and $r$-band data. The profiles fit well with residuals less than $\sim$20 \% in each band.  This suggests that the two component structure we derive from our data is robust.

\begin{table*}[ht]
\caption{Morphological Table of Henize 2-10}
\centering
\begin{tabular}{||c|c|c|c|c||c|c|c|c|c||}
\hline \hline
\multicolumn{5}{|c|}{Double-S\'ersic Fits}&\multicolumn{5}{|c|}{Integrated Data}\\
\hline
Filter &$(r_{\text{eff1}},r_{\text{eff2}})$&$(r_{\text{eff1}},r_{\text{eff2}})$&$(\mu_{e1},\mu_{e2})$ &$(n_1,n_2)$&$\Upsilon$&   $M$   &  L &   Mass   & $\Upsilon_{\text{ave}}$   \\
       &   ($\arcsec$)  &     (pc)       &{ [mag/$\arcsec^2$] }&          & (Outer Comp.)  &{ [mag] }&($\times10^{9} L_{\odot}$)&($\times10^{9} M_{\odot}$)  &     \\
   (1) &     (2)        &        (3)      &          (4)     &  (5)     &   (6)    & (7)   &           (8)                &        (9)                    & (10)               \\
       &                &                 &                  &          &          &       &                              &                               &                    \\
\hline
       &                &                 &                  &          &          &       &                              &                               &                    \\
g      &$6.04 \pm 0.07$ &  259 $\pm$ 3    & 18.7$\pm$0.2     &0.58 $\pm$ 0.01      & 3.5 $\pm$ 0.5 &   -19.2 $\pm$ 0.2    & 5.77 $\pm$ 1.63  & 10.3 $\pm$ 3.9 & 1.8 $\pm$ 0.2  \\
       &$24.72\pm 0.21$ &  1063$\pm$ 9    & 22.0$\pm$0.2     &1.70 $\pm$ 0.01      &               &                      &                  &                &                \\
       &                &                 &                  &                     &               &                      &                  &                &                \\
\hline
       &                &                 &                  &                     &               &                      &                  &                &                \\
r      &5.97 $\pm$ 0.07 &  257 $\pm$ 3    & 19.3$\pm$0.3     & 0.57 $\pm$ 0.01     & 2.5 $\pm$ 0.4 &   -19.8 $\pm$ 0.4    & 6.25 $\pm$ 1.78  & 9.8 $\pm$ 3.0 & 1.6 $\pm$ 0.2   \\
       &24.63$\pm$ 0.20 &  1060 $\pm$ 9   & 22.1$\pm$0.2     & 1.80 $\pm$ 0.01     &               &                      &                  &                &                \\
       &                &                 &                  &                     &               &                      &                  &                &                \\
\hline\hline
\end{tabular}
\vspace{0.5mm} 
\tablenotemark{}
\tablecomments{\textsc{Notes:} The morphological characteristics of Henize 2-10 in two ground-based images; column (1): $g$-and $r$-band images; column (2): effective radii in arc second; column (3): effective radii in pc; column (4): effective SB (at the position of effective radii); column (5): S\'ersic indices; column (6): outer mass-to-light ratios; column (7): absolute magnitudes; column (8): total integrated luminosity under the double-S\'ersic profiles of the outer components; column (9): total stellar integrated mass under the double-Sersic profiles of the innner and outer components combined with the mass of the starbursting region (2$\arcsec$); column 10: the average mass-to-light ratios of the whole galaxy from 0$\arcsec$ to 100$\arcsec$ including three components: inner starbursting region (from 0$\arcsec$ to 2$\arcsec$), intermediate and large radii region of the double-Sersic profiles from $2\arcsec$ to $100\arcsec$.}
\label{table:MorphoHe2-10}
\end{table*}

\begin{table*}[ht]
\caption{Mass \& Light Distribution regions Table of Henize 2-10}
\centering
\begin{tabular}{||c|c|c|c|c|c||}
\hline \hline
Radius     &         $ L_{g} $       &         $ L_{r} $       &    $  M_{\text{dyn}} $   &    $   M_{\text{pop},g}$ &     $  M_{\text{pop},r} $  \\
($\arcsec$)&($\times10^{9} L_{\odot}$)&($\times10^{9} M_{\odot}$)&($\times10^{9} M_{\odot}$)&($\times10^{9} M_{\odot}$)&($\times10^{9} M_{\odot}$)  \\
    (1)    &           (2)           &           (3)          &           (4)           &             (5)        &             (6)           \\
           &                         &                        &                         &                        &                           \\
\hline
           &                         &                        &                         &                        &                           \\
0--2       &    1.43 $\pm$ 0.22      &  1.40 $\pm$ 0.24       &  0.27 $\pm$ 0.11        &           --           &               --          \\
           &                         &                        &                         &                        &                           \\
\hline
           &                         &                        &                         &                        &                           \\
2--6       & 0.19 $\pm$ 0.03         & 0.21 $\pm$ 0.03        & 0.37 $\pm$ 0.15         & 0.36 $\pm$ 0.12        & 0.35 $\pm$ 0.14           \\
           &                         &                        &                         &                        &                           \\
\hline
           &                         &                        &                         &                        &                           \\
6--20      & 0.65 $\pm$ 0.13         & 0.74 $\pm$ 0.17        &           --            & 1.53 $\pm$ 0.13        & 1.50 $\pm$ 0.12           \\
           &                         &                        &                         &                        &                          \\
\hline
           &                         &                        &                         &                        &                           \\
20--100    & 3.46 $\pm$ 1.26         & 3.88 $\pm$ 1.34        &           --            & 8.10 $\pm$ 3.61        & 7.72 $\pm$ 2.54           \\
           &                         &                        &                         &                        &                           \\
\hline\hline
\end{tabular}
\vspace{0.5mm} 
\tablenotemark{}
\tablecomments{\textsc{Notes:} The mass and light distribution regions of Henize 2-10 in two ground-based images; column (1): four distinct regions with different massé and light distributions; column (2): the total intergrated luminosites in each region of $g$-band; column (3): the total intergrated luminosites in each region of $r$-band; column (4): the dynamical mass of each region derived from the linear relation as \citet[\S3.1][]{Marquart2007A&A}; column (5): the stellar population mass of each region in $g$-band; column (6): the stellar population mass of each region in $r$-band. Here the symbol -- means we dot not have the corresponding calculation values for that case. Beyond the radius of 6$\arcsec$ the linear relation as in \citet[\S3.1][]{Marquart2007A&A} is invalid to calculate the stellar dynamical mass, then we only used the stellar population mass.}
\label{table:3partmass}
\end{table*}

\begin{figure}[h]
     \centering
     \epsscale{2.4}
          \plottwo{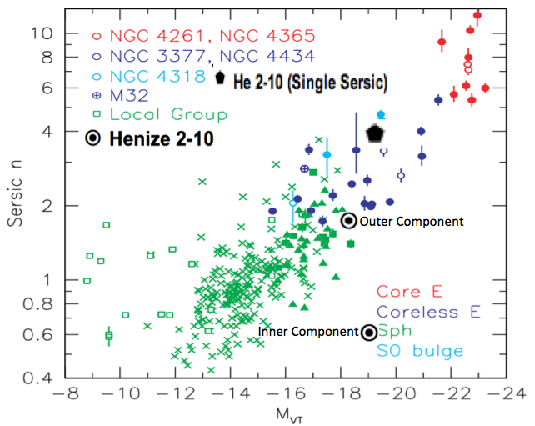}{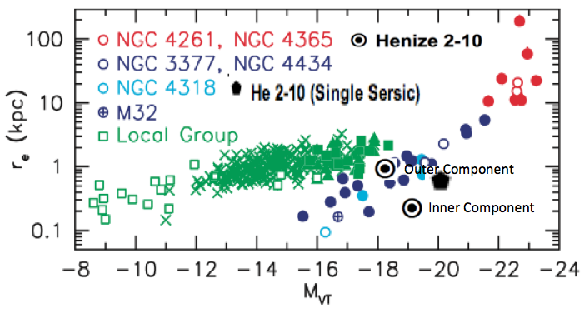}
      \caption[SCP06C1]{ The properties of early-type galaxies as taken from \citet{Kormendy2009ApJ}, Fig. 33 and 37. \textit{Upper Panel:} The top plot shows the correlation between S\'ersic index $n$ and $M_{VT}$. \textit{Bottom Panel:} The correlation of the effective radius, $r_e$, and $M_{VT}$.  The $M_{VT}$ of Henize~2-10 was converted from the absolute maginute in $g$- or $r$-band \citep{Fukugita1996AJ,  Smith2002AJ, Jester2005AJ}. The symbols in both panels are the same: Henize 2-10 is indicated with a bulls-eye, red points are core Es, blue points are extra light Es, green points are Sph galaxies, and turquoise points are S0 bulges as discussed in \citet{Kormendy2009ApJ}. The green triangles show spheroidals from \citet[][]{Ferrarese2006ApJS}. The crosses show spheroidals from \citet[][]{Gavazzi2005A&A}. The open squares are for Local Group spheroidals \citet[][]{Caldwell1999AJ}. The open symbols are non-Virgo Cluster galaxies.}
\label{scalingfig}
\end{figure} 
%%%%%%%%%%%%%%%%%%%%%%%%%%%%%%%%%%%%%%%%%%%%%%%%%%%%%%%%%%%%%%%%%%
\subsection{Mass-to-Light ratio and Mass}
To determine the total mass of Henize 2-10, we split the galaxy into multiple components; for the inner component we derived the mass using kinematics, while at intermediate and large radii, we determined the mass-to-light ratio $\Upsilon$ from the integrated color of the galaxy.  These results are summarized in Table~\ref{table:3partmass}.  

In the outermost of the galaxy beyond 20$\arcsec$, the color is nearly constant, $(g - r)_0=0.75 \pm 0.05$.  Using the dependence of mass-to-light ratio, $\Upsilon$, on the color as describing in \citet[][]{Zibetti2009MNRASZ}, we calculated $\Upsilon_g = 3.5 \pm 0.4$, $\Upsilon_r = 2.5 \pm 0.4$, respectively at large radii.  The luminosity beyond 20$\arcsec$ is $L_g = (3.5 \pm 1.3)\times10^{9} L_{\odot}$ and $L_r = (3.9 \pm 1.3)\times10^{9} L_{\odot}$ corresponding to a total outer mass of $M_{\star,g} = (8.1 \pm 3.6) \times10^{9} M_{\odot}$ and $M_{\star,r} = (7.7 \pm 2.5) \times10^{9} M_{\odot}$ from the $g$- and $r$-band.  Due to the large number of foreground sources requiring masking, the uncertainty in the sky background is substantial at large radii (see Fig.~\ref{sersicfig}).  This is the dominant source of error in our luminosity estimate and these errors are included in the luminosity and mass uncertainties listed in Tables~\ref{table:MorphoHe2-10} and \ref{table:3partmass}.  

At intermediate radii, between 6$\arcsec$ and 20$\arcsec$, the color and thus the mass-to-light ratio varies significantly.  The annulus colors, $(g-r)_0$, are converted into annulus mass-to-light ratios using the \citet[][]{Zibetti2009MNRASZ} relations. These vary between 1.0 and 2.5 in $r$-band and vary between 1.0 and 3.5 in $g$-band.  The total mass between 2$\arcsec$ and 20$\arcsec$ are $\sim (1.89 \pm 0.25) \times10^{9} M_{\odot}$ and $\sim (1.85 \pm 0.26) \times10^{9} M_{\odot}$ for $g$- and $r$-band. The error bars on these data points include the uncertainty on integrated luminosity within this annulus and the scatter $(g-r)_0$, which are converted into the mass-to-light ratio uncertainties \citep[][]{Zibetti2009MNRASZ}.

 Between radii of $2\arcsec$ and $6\arcsec$, we can estimate masses using both the photometric method we used at larger radii, and kinematic estimates.  We estimate the virial mass based on the dispersion in the inner $6\arcsec$ of $45\pm4$~km~$\text{s}^{-1}$ measured by \citet{Marquart2007A&A} and their virial mass estimator.  We derive a mass $(3.7 \pm 1.5) \times 10^8 M_{\odot}$ between $2\arcsec$ and $6\arcsec$.  This dynamical mass agrees well with the photometric mass estimates in this same annulus (Table~\ref{table:3partmass}); for our final calculations we use the $r$-band based photometric mass within this annulus.

Because of the lack of resolution in our Megacam images, the inner color and SB profile within $2\arcsec$ from the center of Henize 2-10 are difficult to determine.  Therefore, we obtained an estimation of the mass of the inner starburst component from kinematics.   Our Gemini/NIFS kinematics show the constant 45~km~$\text{s}^{-1}$ dispersion seen by \citet[][]{Marquart2007A&A} continues to within the central $2\arcsec$.  This corresponds to a dynamical mass of $(2.7 \pm 1.1) \times 10^8 M_{\odot}$  within 2$\arcsec$, which we can add to our photometric estimates from larger radii.  The $r$-band luminosity of this region is $L_{r,2\arcsec}=(1.40 \pm 0.24) \times 10^9 L_{\odot}$, and therefore the mass-to-light ratio in this region drops to $\Upsilon_{r,2\arcsec} = (0.19 \pm 0.08)$  as we might expect for a young population.  

Summing the different radial components in Table~\ref{table:3partmass} together, Henize~2-10 has a total luminosity within 100$\arcsec$ (4.3 kpc) of $L_{r}=(6.3 \pm 1.8) \times 10^{9} L_{\odot}$ ($M_r = -19.8 \pm 0.4$).  The total stellar mass in this aperture is $M_{\star,r}= (9.8 \pm 3.0) \times 10^{9} M_{\odot}$, corresponding to an average mass-to-light ratio of $\Upsilon_r= (1.6 \pm 0.4)$.  Fig.~\ref{m2l_massfig} shows the resulting mass and mass-to-light ratio profile of the galaxy in the $r$-band, $\Upsilon_r$ including the dominant error from the sky determination.  The 100$\arcsec$ aperture may not enclose all the galaxy's light and mass; integrating our SB profile fits to infinity yields $M_r = -19.9$ or about 10\% more light than within our aperture.  The mass derived from $g$-band is consistent within the errors with the $r$-band mass.  Due to the reduced effects of extinction and the good match of our $r$-band profile to the \citet{Micheva2013A&A} profiles, we quote final numbers on the mass based on the $r$-band data.  We note that while the inner starburst region ($r < 2\arcsec$) contributes about 23\% of Henize~2-10's $r$-band luminosity, it contributes about $\sim$ 3\% to the total stellar mass of the galaxy.  

A dynamical estimate of Henize 2-10's total mass within 2.1~kpc was derived from HI kinematics to be 2.7$\times 10^9/{\rm sin}^2\,i\;M_\odot$ by \citet{Kobulnicky1995AJ}.  Our mass estimate within this same aperture is 6.4$\times 10^9$~M$_\odot$ from $r$-band, thus implying inclinations of $i\sim$38$^\circ$.  This translates into an expected ellipticity of $e \sim 0.2$ assuming a thin disk. This is consistent with the ellipticity of 0.15 of the outer isophotes in our ellipse fits and that of \citet{Micheva2013A&A}, suggesting that the gas disk is aligned with the outer stellar component.  

Other previous photometric mass estimates of Henize~2-10 \citep{Kormendy2013ARA&A,Reines2011} have suggested a lower mass for Henize~2-10 $\sim1.4-3.7\times10^9 M_{\odot }$ with factor of 3 uncertainties.  These estimates are based on K-band absolute magnitudes derived from \citet{Noeske2003A&A} $M_{Ks} = -20.86$ over an unspecified aperture and from 2MASS $M_{Ks} = -20.81$ \citep{Skrutskie2006AJ}.  These translate to a $K_s$-band luminosity of $\sim5\times10^9 L_{\odot }$.  Comparing our integrated magnitudes, the color is $r-K_s \sim 1.1$, bluer than expected for a predominantly old population.  However, the color at radii beyond $\sim 10 \arcsec$ between our data and \citet{Noeske2003A&A} is $r-K_s \sim 2.3$ and the profile appears to get redder in the inner part of the galaxy likely due to hot dust; thus it appears the integrated $K_s$-band luminosities are underestimated.  This may be due to use of a smaller aperture than our data.  We also note that $K_s$ band $M/L_s$ are more uncertain than optical $M/L$s due to the poorly understood contributions of supergiant and AGB stars \citep[e.g.,][]{Zibetti2009MNRASZ,Melbourne2012ApJ}.

Although we cannot separate out the masses of the two best-fit S\'ersic components exactly, we use the $\Upsilon_r$ profile to estimate the mass for both components. For the inner component, the integrated luminosity and the average mass-to-light ratio are $L_{r,\text{in}} = (3.1 \pm 0.3) \times10^9 L_{\odot}$ ($M_{r,\text{in}}\sim-19.1$), thus the inner component mass is $M_{\star,r,\text{in}} = (3.9 \pm 0.6) \times10^9 M_{\odot}$. Similarly, we obtained the integrated luminosity $L_{r,\text{out}} = (2.3 \pm 0.4) \times10^9 L_{\odot}$ ( $M_{r,\text{out}}\sim-18.4$), and the mass $M_{\star,r,\text{out}} = (5.8 \pm 1.2) \times10^9 M_{\odot}$ for the outer component.
%%%%%%%%%%%%%%%%%%%%%%%%%%%%%%%%%%%%%%%%%%%%%%%%%%%%%%%%%%%%%%%%%%
\subsection{Henize~2-10 in context}
Apart from the central starburst component, Henize~2-10 appears to be an early-type galaxy.  This is suggested by (1) its central dispersion dominated kinematics \citep[our Fig.~\ref{COfig} and][]{Marquart2007A&A}, (2) the consistently red color of its outskirts.  We argue below that its outer component S\'ersic index ($n \sim 1.8$) is also consistent with this interpretation.

It is useful to compare the morphology and luminosity of Henize~2-10 to other early-type galaxies to better understand its nature and whether or not the presence of a massive BH is suprising.  To facilitate this comparison, we placed our fitted models on diagrams of early-type galaxies from \citet{Kormendy2009ApJ}.  We plotted each component of our double-S\'ersic fits on the plot of absolute magnitude vs. S\'ersic $n$ and effective radius; the inner component is clearly an outlier component in both plots, but the outer component lies at the boundary between what Kormendy considers spheroidal and elliptical galaxies.  One possible interpretation of this is that Henize~2-10 was a typical early-type galaxy before some event formed the nuclear component.  If this is the case, the stellar populations even outside the nucleus should be somewhat younger than the surrounding envelope. In the future, this may result in an age gradient similar to that seen in M32 \citep{Worthey2004AJ} and other lower mass early-type galaxies \citep[e.g.,][]{CidFernandes2005MNRAS, Koleva2009MNRAS, Koleva2011MNRAS, Toloba2014ApJ}.   We can also consider the galaxy as a single component (despite a worse fit to the profile), in which case our single-S\'ersic fits place it firmly amongst the elliptical galaxies with a high S\'ersic index.

Assuming our preferred two component model, the outer component is typical of an old early-type galaxy.  The inner component, with its varying $M/L$, likely represents more recent star formation (perhaps the older portion of the current central starburst), and thus its position off the early-type galaxy locus is unsurprising.  The single component fit overpredicts the flux in the center. Given that most of this flux is in a young starburst it is clear that a single $n$=4 S\'ersic profile does not accurately describe the old population or mass profile of the galaxy. The effective radius and S\'ersic index therefore are unlikely to indicate much about its physical origin.

In the context of Henize~2-10 being a relatively massive $\sim10^{10}$ M$_\odot$ early-type galaxy, it is not surprising that Henize~2-10 hosts a BH.  While dynamical evidence for BHs in galaxies of this mass is almost non-existent \citep{McConnell2013ApJ}, AGN in this mass range and lower have been found \citep{Greene2007ApJ, Barth2008AJ, Desroches2009ApJ, Gallo2010ApJ, Reines2013ApJ, Reines2014ApJ}.  Analyzing these observations, \citet{Greene2012NatCo} and \citet{Miller2014} suggest that the occupation fraction of BHs in early-type galaxies around $10^{10}$ M$_\odot$ is high, $\sim$90\%.  This implies that the BH may be a pre-existing object, and not a result of the intense starburst taking place in the galaxy.

Finally, Henize~2-10 appears to be very isolated, with no known companions within $v < 2000$~km $\text{s}^{-1}$ and 2 degrees ($\sim$~0.3~Mpc) of the galaxy.  Thus the most plausible explanation for Henize~2-10's relatively steep S\'ersic index and current starburst is that it is a late stage merger of two other smaller galaxies as suggested by \citet{Kobulnicky1995AJ} and \citet{Marquart2007A&A}.
%%%%%%%%%%%%%%%%%%%%%%%%%%%%%%%%%%%%%%%%%%%%%%%%%%%%%%%%%%%%%%%%%%%% 
% DYNAMICAL FRICTION OF SSCs SECTION
%%%%%%%%%%%%%%%%%%%%%%%%%%%%%%%%%%%%%%%%%%%%%%%%%%%%%%%%%%%%%%%%%%%% 
\section{The Fate of the Nucleus}
Henize~2-10 contains several super star clusters (SSCs) in the central starburst region \citep{Kobulnicky1995AJ, Mendez1999A&A, Johnson2000AJ, Vacca2002AJ, Chandar2003} and a massive BH with approximative mass of $\sim10^6M_{\odot}$ (with an uncertainty of a factor of $\sim10$) at its center \citep{Reines2011}. We expect the BH and SSCs' orbits to decay due to the gravitational interaction with the surrounding galaxy.  There are two popular models of NSC formation: (1)  dissipative model where the NSC forms {\em in situ} from gas accreted into the center and (2) the dissipationless model where migration of massive star clusters to the center forms the NSC \citep{Chandrasekhar1943, Tremaine1975, Merritt2004, Just2010, Antonini2013}.  

The lack of an existing NSC combined with the presence of a large number of massive young clusters suggests that Henize~2-10 may be an example of ongoing dissipationless nuclei formation.  We examine the expected dynamical friction timescale of this merging in this section based on models by \citet[][]{Antonini2013}.  Assuming that SSCs have King density profiles, orbit within a galaxy with a power-law profile $\rho(r)=\rho_0(r/r_0)^{-\gamma}$, and accounting for mass loss due to the galactic tidal field, the dynamical friction timescale for SSCs to reach the center of the galaxy from dynamical friction mechanism is \citep[Eq. 32 of][]{Antonini2013}:
\begin{equation}
     \tau_{\star}=3\times10^{10}\text{yr}\frac{(4-\gamma)\sqrt{\gamma}\ln\Lambda_3^{-1}}{(3-\gamma)^3F(\gamma)}\rho_{0,5}r_{0,700}^3\sigma_{K,10}^{-3}\Big(\frac{r_{\text{in}}}{r_0}\Big)^{3-\gamma}
\end{equation}
with the central density, $\rho_{0,5}=\rho_0/5 M_{\odot}\text{pc}^{-3}$, the scale radius, $r_{0,700}=r_0/700\,\text{pc}$, the Coulomb logarithm, $\ln\Lambda_3=\ln\Lambda/3$, the cluster mass, $m_{cl,6}=m_{cl}/10^6M_{\odot}$, the central dispersion velocity of SSC, $\sigma_{K,10}=\sigma_K/10$~km~$\text{s}^{-1}$, and the initial distance, $r_{\text{in}}$, of the SSC from the galactic center. The $\gamma$ dependent coefficient, $F(\gamma)=(0.193,0.302,0.427)$ for $\gamma=(1, 1.5, 2)$ corresponding to the analytical formula can be found in \citep{Merritt2004}.
\begin{equation}
     F(\gamma)=\frac{2^{\beta}\Gamma(\beta)(2-\gamma)^{-\gamma/(2-\gamma)}}{\sqrt{2\pi}\Gamma{(\beta-3/2)}}\int_0^1y^{1/2}\Big(y+\frac{2}{2-\gamma}\Big)dy
\end{equation}
where $\beta=(6-\gamma)/2(2-\gamma)$. $\ln\Lambda$ has a value from 2 to 7 \citep{Agarwal2011}.

We note that the equation (1) does not take into account the presence of a BH; however, given the low mass ($\sim 10^6$~$M_\odot$) and small sphere of influence (a few pc) of the BH, clusters should reach a radius similar to the size of typical NSCs before feeling the effects of the BH. 

We also caution that in some cases, dynamical friction appears to be less efficient than predicted from theory.  Most notably, \citet{Lotz2001ApJ} suggest that the faintness of nuclei in low mass dE galaxies ($M_V > -14$) indicates inefficient dynamical friction, perhaps due to supernova feedback or mismatches of the dark matter halo with the visible baryonic component.  Given the much higher luminosity of Henize~2-10 and the likely dominance of baryonic matter near the center of the galaxy, it is in a regime where \citet{Lotz2001ApJ} find evidence for dynamical friction working as expected. 

We apply these dynamical friction timescale results to explore the possibility of forming a NSC at the center of Henize~2-10 due to the migration of SSCs. Before we can do this, we need to determine several parameters for the galaxy and clusters, which we do in the following subsections.  This includes a determination of the cluster galactocentric positions, profiles, and masses, as well as the density profile of the galaxy.  We derive these properties using a combination of high resolution {\it ACS/HRC} imaging and Gemini/NIFS spectroscopy. 
%%%%%%%%%%%%%%%%%%%%%%%%%%%%%%%%%%%%%%%%%%%%%%%%%%%%%%%%%%%%%%%%%%%%%%%%%%%%%
\subsection{Determination of Cluster photometric parameters, description of the cluster sample}
We identified the brightest 11 SSCs in the nuclear region using the {\it HST} images as shown in Fig.~\ref{SSCs}.  These clusters are also seen in the Gemini/NIFS observations, but are not as well resolved; the brightest four clusters are clearly visible in both data sets and allow us to astrometrically align our Gemini/NIFS image with the {\it HST} data.  

Four of the SSCs identified here were previously identified by \citet[][]{Chandar2003}; we label these CLTC1-4.  The brightest SSC is CLTC1, and this cluster is shown at the center of our maps.  Seven other bright SSCs are also identified in the {\it HST} imaging, and we number these 5 through 11.  The coordinates of all clusters are given in Table~\ref{table:CLTC1-4}.  For purposes of our dynamical friction timescales calculation, we assumed the galactocentric radius of the clusters is their projected distance to the BH.  Given the compact region over which clusters are found, it is likely that this distance is within $\sim$50\% of their true galactocentric distance.

For all 11 clusters, we performed King model fits to the {\it HST ACS/HRC} $F814W$ images.  We followed the procedure described in \citet[][]{Larsen2001} assuming concentration parameter, $c=30$.  We fit the data with the {\tt iSHAPE} rountine which is part of the {\tt BAOLAB} software \citep{Larsen1999}. To fit each SSC's size (the core radius or effective radius) in the image, {\tt iSHAPE} needs a corresponding PSF, and a charge diffusion kernel image.   The PSF is generated by the {\tt Tiny Tim} routine \citep{Krist2011} and oversampled by a factor of 10 for use with {\tt iSHAPE}. The charge diffusion kernel is generated from header file of distorted PSF image when we ran the $\tt{tiny3}$ task of {\tt Tiny Tim} routine. The core radius, $r_c$, and effective radius, $r_e$, from the best fit King model are shown in column 4 and 5 of Table~\ref{table:CLTC1-4}.  Integrated fluxes of the clusters in both $F814W$ and $F205W$ are calculated using aperture photometry (see \S4.2).   The luminosity in $F814W$ is given in column 6, while the color $F814W-F205W$ in VEGA system of integrated fluxes of the two bands is given in column 9.  This color is useful for evaluating the similarities in stellar population and reddening among the clusters (see \S4.3).  

\begin{table*}[ht]
\caption{Dynamical Quantities for SSCs}
\centering
\begin{tabular}{ccccccccc}
\hline \hline
SSCs&$\sigma_{\text{cen}}$ &$\sigma_{\text{aper}}$&$r_{\text{eff}}$&$r_c$&$L_{F814W}$&$M$&$\Upsilon_{F814W}$& $F814W-F205W$\\
(1) &            (2)     &       (3)          &      (4)      &(5) &    (6)   & (7)&         (8)     & (9)\\
    &(km $\text{s}^{-1}$) &(km $\text{s}^{-1}$)&      (pc)     &(pc)&$(\times10^7 L_{\odot})$&$(\times10^6 M_{\odot})$& &  \\
\hline     
      &                  &                  &                 &                 &                 &                 &    &    \\      
CLTC1 $08^h36^m15^s.199$ &                  &                 &                 &                 &                 &    &    \\ 
  $-26^{\circ}24^{\prime}33^{\prime\prime}.62$ & 17.6 $\pm$ 1.6   & 20.1 $\pm$ 0.6 & 3.43 $\pm$ 0.07 & 1.27 $\pm$ 0.03 & 3.59 $\pm$ 0.93 & 2.30 $\pm$ 0.60 & 0.06 $\pm$ 0.02 & 2.2\\
      &                  &                  &                 &                 &                 &                 &    &   \\
\hline
      &                  &                  &                 &                 &                 &                 &    &    \\
CLTC2 $08^h36^m15^s.164$ &                  &                 &                 &                 &                 &    &    \\ 
  $-26^{\circ}24^{\prime}33^{\prime\prime}.71$  & 13.8 $\pm$ 1.2   &                & 2.07 $\pm$ 0.03 & 0.77 $\pm$ 0.01 & 1.45 $\pm$ 0.56 & 0.92 $\pm$ 0.61 &  &  2.0\\
      &                  &                  &                &                 &                 &                 &    &    \\
CLTC3 $08^h36^m15^s.134$ &                  &                &                 &                 &                 &    &    \\ 
  $-26^{\circ}24^{\prime}34^{\prime\prime}.73$  & 15.4 $\pm$ 1.3   &                & 2.07 $\pm$ 0.09 & 1.00 $\pm$ 0.02 & 1.78 $\pm$ 0.74 & 1.14 $\pm$ 0.78 &    &  2.1\\
      &                  &                  &                 &                 &                 &                 &    &    \\
CLTC4 $08^h36^m15^s.113$ &                  &                 &                 &                 &                 &    &    \\ 
  $-26^{\circ}24^{\prime}34^{\prime\prime}.80$  & 14.4 $\pm$ 1.1   &                 & 1.89 $\pm$ 0.11 & 0.71 $\pm$ 0.05 & 1.42 $\pm$ 0.53 & 0.91 $\pm$ 0.60 &    &  1.9\\
      &                  &                  &                 &                 &                 &                 &    &    \\
\hline
      &                  &                  &                 &                 &                 &                 &    &    \\
5 $08^h36^m15^s.245$     &                  &                 &                                   &                 &    &     \\
  $-26^{\circ}24^{\prime}34^{\prime\prime}.81$  &11.0 $\pm$ 1.0   &                 & 1.42 $\pm$ 0.13 & 0.53 $\pm$ 0.07 & 0.63$\pm$ 0.12 & 0.40 $\pm$ 0.21 &    &1.7\\
      &                  &                  &                 &                 &                 &                 &    &    \\
6 $08^h36^m15^s.234$     &                  &                 &                                   &                 &    &     \\
  $-26^{\circ}24^{\prime}34^{\prime\prime}.10$  &11.0 $\pm$ 1.5   &                 & 1.42 $\pm$ 0.17 & 0.53 $\pm$ 0.04 & 0.63 $\pm$ 0.12 & 0.40 $\pm$ 0.21 &    &1.8\\
      &                  &                  &                 &                 &                 &                 &    &    \\
7 $08^h36^m15^s.236$     &                  &                 &                                   &                 &    &     \\
  $-26^{\circ}24^{\prime}34^{\prime\prime}.40$  &9.1 $\pm$ 1.1    &                 & 2.38 $\pm$ 0.34 & 0.89 $\pm$ 0.21 & 0.72 $\pm$ 0.12 & 0.46 $\pm$ 0.25 &    &2.0\\
      &                  &                  &                 &                 &                 &                 &    &    \\
8 $08^h36^m15^s.231$     &                  &                 &                                   &                 &    &     \\
  $-26^{\circ}24^{\prime}34^{\prime\prime}.63$  &12.5 $\pm$ 0.8   &                 & 1.24 $\pm$ 0.04 & 0.46 $\pm$ 0.02 & 0.70 $\pm$ 0.12 & 0.45 $\pm$ 0.24 &    &1.8\\
      &                  &                  &                 &                 &                 &                 &    &    \\
9 $08^h36^m15^s.225$     &                  &                 &                                   &                 &    &     \\
  $-26^{\circ}24^{\prime}34^{\prime\prime}.26$  &9.3 $\pm$ 0.6    &                 & 1.00 $\pm$ 0.03 & 0.37 $\pm$ 0.01 & 0.32 $\pm$ 0.07 & 0.20 $\pm$ 0.11 &    &2.1\\
      &                  &                  &                 &                 &                 &                 &    &    \\
10 $08^h36^m15^s.216$    &                  &                 &                                   &                 &    &     \\
   $-26^{\circ}24^{\prime}34^{\prime\prime}.76$ &11.6 $\pm$ 1.3   &                 & 1.45 $\pm$ 0.05 & 0.46 $\pm$ 0.02 & 0.70 $\pm$ 0.12 & 0.45 $\pm$ 0.24 &    &2.0\\
      &                  &                  &                 &                 &                 &                 &    &    \\
11 $08^h36^m15^s.173$    &                  &                 &                                   &                 &    &     \\
  $-26^{\circ}24^{\prime}34^{\prime\prime}.25$  &6.2 $\pm$ 0.9    &                 & 2.21 $\pm$ 0.07 & 0.37 $\pm$ 0.01 & 0.32 $\pm$ 0.07 & 0.20 $\pm$ 0.11 &    &1.6\\
      &                  &                  &                 &                 &                 &                 &    &    \\
\hline
\vspace{0.3mm} 
\end
{tabular}
\tablenotemark{}
\tablecomments{\textsc{Notes:} Individual cluster data for SSCs identified in Fig.~\ref{SSCs}.  Column (1): Identified, CLTC1-4 corresponds to clusters identified by \citet[][]{Chandar2003}, the rest are identified here with coordinates given, column (2): Velocity dispersion from central pixel; column (3): Velocitiy dispersion in 0$\farcs$2 radius; column (4): effective (half-light) radius; column (5): core radius; column (6): $F814W$ Luminosity; column (7): Mass estimate; column (8): mass-to-light ratio; column (9): $F814W-F205W$ color. A dynamical mass and $\Upsilon_{F814W}$  were measured only for CLTC1, all other clusters assumed the same $\Upsilon_{F814W}$ to derive their mass.}
\label{table:CLTC1-4}
\end{table*}
%%%%%%%%%%%%%%%%%%%%%%%%%%%%%%%%%%%%%%%%%%%%%%%%
\subsection{Virial SSC mass determination}
\citet[][]{Chandar2003} estimated masses of the 4 brightest SSCs, CLTC1-4, using stellar population model fits of ultraviolet and blue {\it HST/STIS} spectra, finding a mass for CLCT1 of $\sim$ 4.5 $\times$10$^5$~M$_\odot$.  However, these fits have significant uncertainties due to dust extinction.  Here, we measured the dispersion of the brightest cluster, CLTC1, from our Gemini/NIFS data and used it to determine a more robust virial mass estimation \citep[e.g.,][]{Peterson1975, Larsen2002, McCrady2007, Strader2009}.  We could only determine the internal dispersion of CLTC1 because of galaxy contamination in the other fainter clusters.
  
\begin{figure*}[ht]
     \centering
      \epsscale{1.17}
          \plotone{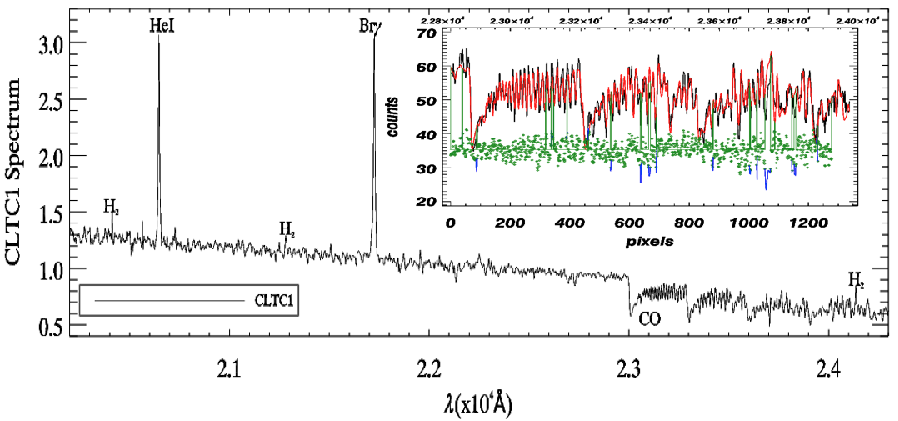}
     \caption[SCP06C1]{The normalized spectra of the brightest SSC (CLTC1) after subtracting off the local background.  Labels show emission from $\text{Br}\gamma$, HeI, and $H_2$ (1-0 S(1) and 1-0 Q(1)), as well as the CO absorption bandheads.  The inset plot shows the kinematic fit to the CO absorption bandhead region. The black solid line is the observed spectrum, the solid red line is the best fit kinematic model. The green data points are the data-model residuals; blue lines indicate the regions of the spectrum excluded from the fit}.
\label{SSC1st}
\end{figure*}

Fig.~\ref{SSC1st} shows the full spectrum and CO absorption bandhead of CLTC1 from Gemini/NIFS data after subtracting an estimation of the background light of the galaxy.  We estimated the galaxy background value by taking the median value of pixels surrounding CLTC1 but avoiding contamination from neighboring clusters. For our Gemini/NIFS data of the SSCs, only the spectrum of CLTC1 was cleanly separated from the surrounding galaxy. Specifically, using annuli surrounding the clusters, we find that the background contamination in CLTC1 is $\sim$20\%, while in the rest of the clusters it is $>$45\%.    

As for the galaxy as a whole, we fit the kinematics in the CO absorption bandhead to estimate velocity dispersion of CLTC1 using PPXF (see \S2.4).  As the cluster is nearly unresolved in our data, we sum the light within an 0$\farcs$2 radius aperture over which the source is well above the galaxy background.  Our resulting best-fit dispersion for this spectrum is $\sigma = 20.1 \pm 0.6$~km~$\text{s}^{-1}$.  Given that this aperture extends out $\sim$2$\times$ the effective radius of the cluster, we assume that this dispersion is reflective of the cluster's global velocity dispersion $\sigma_{\infty}$, which is used in virial mass estimation.  To assess the robustness of this dispersion, we conducted several tests: (1) We determined the dispersion value in just the central pixel of the cluster to be $17.6 \pm 1.6$~km~$\text{s}^{-1}$.  (2) We varied the pixels used for background subtraction and find dispersion results that vary from 17 to 21~km~$\text{s}^{-1}$. (3) We fit the data to no dispersion model and find that the resulting reduced $\chi^2$ is higher than the best-fit model by 20\%.  These tests suggest that our dispersion value is clearly resolved by the data, but that the systematic error is more like $\sim$3~km~$\text{s}^{-1}$.  The cluster's radial velocity is blue-shifted by 6~km~$\text{s}^{-1}$ relative to the systemic velocity of the galaxy.  

The virial mass is calculated using \citet[][]{Strader2009}:
\begin{equation}
     M_{\text{vir}}=\frac{7.5\sigma_{\infty}^2r_{\text{hm}}}{G}=4\times a\frac{\sigma_{\infty}^2r_{e}}{G}
\end{equation} 
where $\sigma_{\infty}$ is the global velocity dispersion, $r_{\text{hm}}$ is the half mass radius, and $r_{\text{hm}}=4r_e/3$, $r_e$ is the effective radius or half-light radius. The structural parameter, $a$, of globular clusters (GCs) is 2.5 \citep[][]{Spitzer1987}. We determined $r_{\text{hm}}$ or $r_{e}$ from King model fits as discussed above. Based on the measured dispersion of CLTC1, we estimated a mass of 2.3~$\times 10^6 M_{\odot}$.  Assuming a 3~km~$\text{s}^{-1}$ error on the dispersion, this translates to a $\sim28$\% error (0.6$\times 10^6 M_{\odot}$) on the mass.

For CLTC1, \citet[][]{Chandar2003} found a mass of 4.5~$\times 10^5 M_{\odot}$ from their spectral fitting, assuming a Starburst 99 \citep{Leitherer1999ApJ} model with a Salpeter initial mass function (IMF) and with lower and upper masses of 1 $M_{\odot}$ and 100 $M_{\odot}$, respectively.  This mass is a factor of 5$\times$ lower than our dynamical mass estimate of 2.3~$\times 10^6 M_{\odot}$. This discrepancy could be due to the assumption by \citet[][]{Chandar2003} of a 1~M$_\odot$ lower limit on the IMF mass in their models which will lead to an order of magnitude underestimate in the total stellar mass relative to a Kroupa IMF \citep{Kroupa2001MNRAS}.
%%%%%%%%%%%%%%%%%%%%%%%%%%%%%%%%%%%%%%%%%%%%%%%%%%%%%%%%%%%%%%%%%%%%%%%%%%%%%
\subsection{Masses of other clusters}
In this section, we made rough estimates of the masses of the other optically visible clusters in the nuclear region using our derived M/L for cluster CLTC1.  While this assumption is an undoubtedly oversimplification, it is a reasonable one for two reasons.  First, the star clusters studied in this region have all been found to have ages $<$~5~Myr \citep{Chandar2003}, as might be expected from the prodigious star formation rate of $1.9 M_{\odot}\text{yr}^{-1}$ \citep{Mendez1999A&A, Engelbracht2005ApJ, Calzetti2007ApJ}.  Second, our measurements of the color ($F814W-F205W$) between $F814W$ and $F205W$ bands (Table~\ref{table:CLTC1-4}) in most of the clusters suggests they have similar ages and extinctions because the colors change slightly over the NIFS field of view (FOV) and {\it HST} data.  We noted that colors were calculated from a fixed $0\farcs2$ aperture in each cluster; no correction was made for the differing PSF between the bands; however, the difference in encircled energies for a point source is relatively small (75\% in {\it F814W}, 65\% for {\it F205W}), and likely quite similar for all clusters.  

The clusters' luminosities used here are the integrated King model luminosities which are obtained from the {\tt iSHAPE} rountine.  Each cluster's luminosity is ouput as a parameter, we then correct for Galactic extinction to get the total luminosity shown in column 6 of Table~\ref{table:CLTC1-4}. The mass-to-light ratio of CLTC1 in $F814W$ estimated using the virial mass estimate from \S4.2 gives a $\Upsilon_{F814W} \sim 0.064$. This value is quite consistent with the mass-to-light ratio in H band of 0.05 found for the larger starburst region by \citet[][]{Marquart2007A&A}. Assuming this constant mass-to-light ratio for all the SSCs, we derived masses as given in Table~\ref{table:CLTC1-4}.  

We found a total mass in young clusters in the nuclear region to be $\sim$7.4~$\times 10^6 M_{\odot}$. This nuclear region overlaps with region A from \citet[][]{Johnson2000AJ}, who derived a total mass between $1.6-2.6\times 10^6 M_{\odot}$ from UV spectra.  This large discrepancy might arise from their assumed IMF, which like \citet{Chandar2003} used Salpeter IMF between 1-100 $M_\odot$ (see above). 

\begin{table}[ht]
\caption{Dynamical Friction Quantities for SSCs}
\centering
\begin{tabular}{ccccc}
\hline \hline
SSCs & $r_{\text{in}}$    &$r_{\text{lt}}$&$r_{\text{dis}}$&    $\tau_{\rm dyn}$     \\
(1)  &   (2)            &    (3)       &       (4)    &           (5)          \\
     &   (pc)           &     (pc)     &       (pc)   &$\times10^8\text{(yr)}$ \\
\hline
       &                &              &              &                        \\
CLTC1  & 64.8 $\pm$ 1.4 &       7      &     6        &  $1.7 \pm$ 0.5         \\
       &                &              &              &                        \\
\hline
       &                &              &              &                        \\
CLTC2  & 36.9 $\pm$ 1.4 &       5      &     4        &  $0.8 \pm$ 0.2         \\
       &                &              &              &                        \\
CLTC3  & 20.4 $\pm$ 1.4 &       2      &     7        &  $0.14 \pm$ 0.04       \\
       &                &              &              &                        \\
CLTC4  & 11.2 $\pm$ 1.4 &       2      &     3        &  $0.04 \pm$ 0.01       \\
       &                &              &              &                        \\
5      & 86.4 $\pm$ 1.4 &       7      &     2        &  $34 \pm$ 10           \\
       &                &              &              &                        \\
6      & 77.6 $\pm$ 1.4 &       6      &     2        &  $24 \pm$ 7            \\
       &                &              &              &                        \\
7      & 114.5$\pm$ 1.4 &       6      &     6        &  $22 \pm$ 1            \\
       &                &              &              &                        \\
8      & 130.7$\pm$ 1.4 &       7      &     3        &  $11 \pm$ 3            \\
       &                &              &              &                        \\
9      & 74.1 $\pm$ 1.4 &       4      &     2        &  $16 \pm$ 5            \\
       &                &              &              &                        \\
10     & 70.2 $\pm$ 1.4 &       4      &     2        &  $7  \pm$ 2            \\
       &                &              &              &                        \\
11     & 55.8 $\pm$ 1.4 &       2      &     5        &  $24 \pm$ 7            \\
       &                &              &              &                        \\
\hline \hline
\vspace{0.3mm} 
\end{tabular}
\tablenotemark{}
\tablecomments{\textsc{Notes:} Dynamical friction quantities of 11 SSCs in the FOV of NIFS. Column (1): SSCs' labels. Column (2): initial galactocentric position; column (3): limiting tidal radius at the SSCs current position; column (4): distance from the galactic center at which the SSC is disrupted; Column (5): dynamical timescales calculated from Eq. (1) which incorporates this effect of the interaction with the galactic tidal field into the calculation. The numbers in this column assume a density profile with $\gamma$=0.5.}
\label{table:SSC_dynafriction}
\end{table}
%%%%%%%%%%%%%%%%%%%%%%%%%%%%%%%%%%%%%%%%%%%%%%%%%%%%%%%%%%%%%%%%%%%%%%%%%%%%%
\subsection{Dynamical Friction Timescale}
\subsubsection{Galaxy Density Profile Determination}
In addition to the cluster parameters, we require the central density profile of the galaxy to determine the dynamical friction timescales.   Following \citet[][]{Antonini2013}, we characterize the central starburst region of Henize~2-10 with a simple power-law density profile, $\rho(r)=\rho_0(r/r_0)^{-\gamma}$.  There are three parameters we need to specify: the central stellar mass density, $\rho_0$, the scale radius of the bulge, $r_0$, and power-law index, $\gamma$. 

Due to the presence of the central starburst, measuring the central profile of the galaxy is not straightforward.  While our SB profile fits did not include the central region, we can extrapolate the expected density from those profiles to determine the central density profile.  We are thus assuming the central density is dominated by the extrapolation of the older stellar populations as captured by our two-S\'ersic component fit, and not the young starburst which likely contributes only a small fraction of the mass despite dominating the optical luminosity at the center. 

To facilitate deprojection of the SB profile to a 3D mass density profile, we create a Multiple Gaussian Expansion (MGE) \citep{Emsellem1994A&A, Cappellari2002MNRAS} model based on the two component SB profile presented in \S3.  In order to translate this luminosity profile into a mass profile, we use the observed kinematics in the central region \citep[$\sigma = 45$~km~$\text{s}^{-1}$][]{Marquart2007A&A} to constrain the mass-to-light ratio.  Specifically, we created a Jeans Axisymmetric Model \citep{Cappellari2008MNRAS} to predict the dispersion in the central region based on the MGE model.  We fix the mass-to-light ratio of the outer component's S\'ersic profile to $M/L_r = 2.5$, while varying that of the inner S\'ersic component to match the observed dispersion.  The best-fit requires a $M/L \sim 2$ for this inner component. We then fit the deprojected MGE density model to the $\rho(r)=\rho_0(r/r_0)^{-\gamma}$ profile within the central 3$\arcsec$. The best fit parameters are: $\rho_0 = 78M_{\odot}\;\text{pc}^{-3}$, $r_0 = 95$~pc, and $\gamma = 0.5$.

\subsubsection{Cluster Limiting Radii \& Dissolution}
Now that we have a model for the central potential in the galaxy, we can determine whether each cluster is currently tidally limited and whether the clusters will reach the center without being disrupted by the gravitational field of the galaxy and BH.  Specifically, we use \citet[Eq. 23 of][]{Antonini2013} to calculate each cluster limiting tidal radius $r_{lt}$: 
\begin{equation}
     r_{lt}=\frac{\sigma_K}{\sqrt{2}}\Big[4\pi G\rho_0\Big(\frac{r}{r_0}\Big)^{-\gamma}\frac{\gamma}{3-\gamma}+\frac{3GM_{\bullet}}{r^3}\Big]^{-1/2}
\end{equation}
 where $r$ is the projected distance between each cluster and the position of the BH, and $M_{\bullet}$ is the BH's mass.  The projected distance of each SSC is listed in the column 2 of Table~\ref{table:SSC_dynafriction}. $\sigma_K$ is SSC's velocity dispersion which is estimated from the mass and profile estimate listed in the column 2 of Table~\ref{table:MorphoHe2-10}, except for CLTC1 where we use our measurement. We find that in each cluster the limiting radius is larger than the best-fit core radius, but smaller than the tidal radius (we fit all data to a $c=30$ King model, thus $r_t = 30 r_c$ listed in column 3 of Table~\ref{table:SSC_dynafriction}). This means each cluster is already being limited by the gravitational potential of the galaxy, and thus should have a dynamical friction timescale given by equation 1 which incorporates the affects of mass loss as the cluster moves inwards. 

The clusters may be disrupted by the external gravitational field before reaching the center of the galaxy; following \citet{Antonini2013} we assume this occurs when the limiting radius is equal to the core radius of each cluster.  This galactocentric radius is given in column 4 of Table \ref{table:SSC_dynafriction}; each cluster should survive until it is within the central 10~pc (and in most cases the central 5~pc).  Given that typical NSCs have effective radii of $\sim$4~pc \citep[e.g.,][]{Cote2006ApJ}, each cluster is likely to add at least some of its mass to a forming NSC if their dynamical friction timescales are short enough. 

Internal processes could also shorten the lifetimes of these clusters; substantial natal gas or stellar wind mass loss could cause the clusters to dissolve.  However, the clusters are already optically visible and thus have likely lost their natal gas.  Furthermore, stellar wind mass loss represents $<$30\% of the clusters mass within the first 500 Myr of a cluster \citep[][]{Gieles2010IAUS}, and thus this effect is unlikely to dissolve these massive clusters.  Finally, we expect dissolution from two body relaxation to be very slow; the clusters have relaxation times of $>5\times10^8$~yr; dissolution due to dynamical evolution occurs over many relaxation times \citep[][]{Spitzer1987, Binney2008.book}.

\subsubsection{Dynamical Friction Timescales}
Finally, we calculate the clusters' dynamical fricition timescales using Eq.~(1) \citep[Eq. 32 of][]{Antonini2013}. These timescales are given in column 5 of Table~\ref{table:SSC_dynafriction}; for the most massive clusters these timescales are $<$10$^9$ years.  The most massive cluster, CLTC1 has a timescale of just 170 Myr.  Thus a nuclear cluster well in excess of $10^6$~M$_\odot$ is likely to form within a few hundred Myr.  The end product will have the appearance of a NSC; the faintest NSCs in $\sim10^{10}$~M$_\odot$ galaxies typically have NSCs of $10^6$~M$_\odot$ \citep[][]{Seth2008ApJ}.  Because the timescale of formation is so short relative to the age of the Universe, the state we are seeing in Henize~2-10 should be rare.

Henize~2-10 thus provides an interesting snapshot of NSC formation in action and constraints on the coevolution of NSCs and BHs.  While there have been previous suggestions that massive clusters may provide the seeds for BHs \citep[e.g.,][]{PortegiesZwart2004}, if there is any coevolution between the two components in Henize 2-10 it would be due to merger fed mutual growth as modeled by \citet{Hopkins2010aMNRAS, Hopkins2010bMNRAS}.   Thus while the NSC is forming, the BH may also grow through gas accretion or the accretion of stars \citep{Bromley2012ApJ}.  However, as we argued in \S3.3, it appears likely based on the galaxy's overall mass and early-type nature that a BH was already present in the galaxy, and that the growth of the NSC is occurring after BH formation.  This would suggest that the formation of the NSC and BH, at least in this galaxy, is largely uncorrelated. 
%%%%%%%%%%%%%%%%%%%%%%%%%%%%%%%%%%%%%%%%%%%%%%%%%%%%%%%%%%%%%%%%%%%%%%%%%%%%%
% NUCLEAR SPECTRUM SECTION.
%%%%%%%%%%%%%%%%%%%%%%%%%%%%%%%%%%%%%%%%%%%%%%%%%%%%%%%%%%%%%%%%%%%%%%%%%%%%%
\section{Henize 2-10 Nuclear Spectrum}\label{sec:nuclear spec}
 There are several ways an accreting BH could be detected from our $K$-band NIFS spectrum:  (1) dynamical modeling of the stellar velocity dispersion in the vicinity of BH \citep[e.g.,][]{Krajnovic2009MNRAS}, (2) detection of coronal line emission such as \ion{Al}{9}, which requires extreme emission from the region around the accreting BH to excite \citep[e.g.,][]{Storchi2009MNRAS}, or (3) detection of other emission lines, such as Br$\gamma$, which might be associated with X-ray radiation from the accreting region. 

To determine if the BH may be dynamically detectable from our kinematics observations we first consider the sphere of influence of the BH.  The mass of the BH was estimated to be $2\times10^6$~M$_\odot$ by \citet{Reines2011} based on the fundamental plane which provides order of magnitude estimates of BH masses \citep[e.g.,][]{Merloni2003MNRAS}.  The dispersion in the inner region of the galaxy outside of the star clusters is $\sigma \sim 45$~km~$\text{s}^{-1}$.  Using this dispersion, the sphere of influence (SOI) is given as $r_{\rm SOI} = G M_{\rm BH} / \sigma^2 \sim 4$~pc or $\sim0\farcs1$.  This is comparable to the PSF FWHM of $0\farcs15$ and thus it may be possible to see the effect of a BH on the kinematics. We note however that there is considerable uncertainty in the BH mass estimate and therefore in the sphere of influence.

There is no clear detection of a dispersion peak or enhanced rotation in the vicinity of the BH; in fact, the dipsersion is lower at the location of the BH than in the surrounding areas (Fig.~\ref{COfig}).  To better understand the expected signal from the BH, we use our Jeans model for the center of the galaxy derived in \S4.4 to simulate our NIFS observations with the addition of a BH with mass between 0 and 10$^7$~M$_\odot$.  We find that for a BH of mass $2\times10^6M_{\odot}$ we would expect an increase in dispersion of $\sim 1.5$~km $\text{s}^{-1}$; this is smaller the 1$\sigma$ random errors on our measurement and thus a BH of this mass cannot be ruled out.  We can place a firm upper limit of $\sim10^7M_{\odot}$ on the BH which would increase the dispersion by $\sim$7~km $\text{s}^{-1}$ and thus be clearly detectable at the 3$\sigma$ level in our data.  Another possibility is that the BH is significantly obscured and the stars we are seeing are in the foreground of the BH.  If this were the case, the obscuring material would need to be on scales larger than the sphere of influence.

 We also searched for \ion{Al}{9} coronal line emission in the vicinity of the BH and did not find a convincing detection. This line has an ionization energy of 285 eV, thus requiring X-ray emission likely only generated by an AGN \citep[e.g.,][]{satyapal2008ApJ}.  The only other coronal line in our bandpass is the \ion{Ca}{8} line which falls on top of the CO bandhead region and thus is challenging to detect at weak flux levels.  We place a 3$\sigma$ upper limit on the \ion{Al}{9} luminosity of $4\times 10^{-17}\;\text{erg}$ $\text{s}^{-1}\;\text{cm}^{-2}$.

\begin{figure}[h]     
     \centering
     \epsscale{1.25}
     \plotone{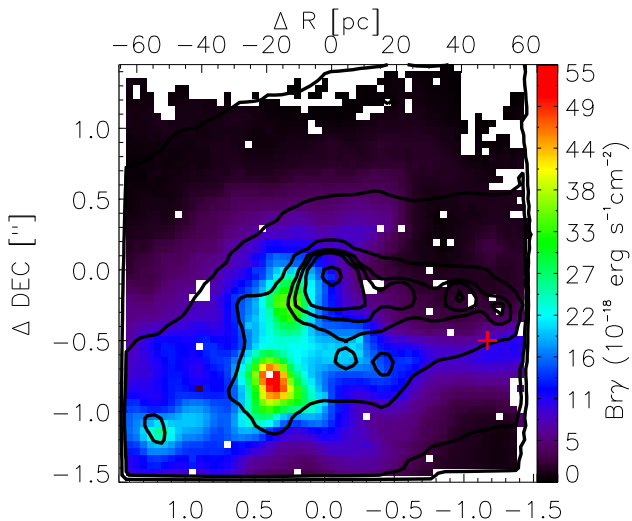}   
     \caption[SCP06C1]{Br$\gamma$ emission of the starburst region in the NIFS $3\arcsec\times3\arcsec$ FOV.  Colors and color bar indicate the flux of the Br$\gamma$ line emission. White areas have Br$\gamma$ emission with SNR $<$ 3.  The red cross represents the position of BH; significant Br$\gamma$ emission is located at this position. Contours indicate $K$-band surface brightness of 15.7, 15.9, 16.1, 16.6, and  17.7 mag/$\arcsec^2$. The coordinates are centered on the continuum peak of the brightest cluster, CLTC1.}
\label{Brgamma}
\end{figure}
  
\begin{figure}[h]
     \centering
     \epsscale{1.15}
          \plotone{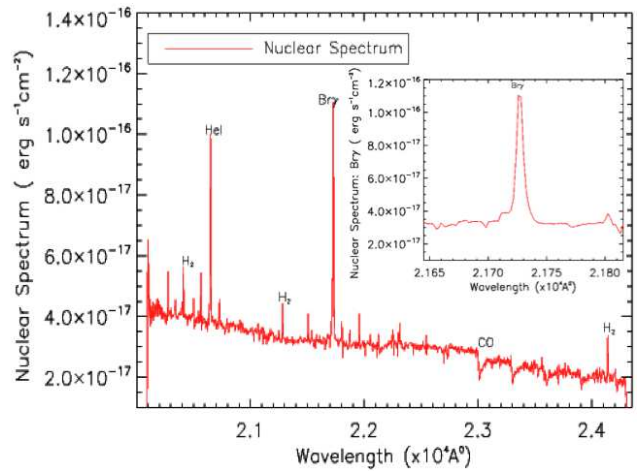}
     \caption[SCP06C1]{NIFS spectrum of AGN within $0.2\arcsec$ aperture with detected emission lines of $\text{Br}\gamma$, HeI, and $H_2$ (1-0 S(1) and 1-0 Q(1)) and CO absorption lines are marked. The smaller panel indicates the significantly detected Br$\gamma$ emission only.}
\label{AGN}
\end{figure}

Fig.~\ref{Brgamma} shows the Br$\gamma$ emission map within our NIFS FOV.  This map shows a clear enhancement of Br$\gamma$ emission at the AGN's position (red cross is the BH's suggested location with an astrometric uncertainty of $\sim$0$\farcs$1).  To examine this Br$\gamma$ peak in more detail, we create a nuclear spectrum around this source with a radius of $0.2\arcsec$ and sky annulus is about 10\% of the total Br$\gamma$ flux.  This spectrum is shown in Fig.~\ref{AGN}; the main panel shows the spectrum, while the small upper right panel shows a zoom-in of the dominant emission line, Br$\gamma$.  No broad Br$\gamma$ emission is seen.  

The total Br$\gamma$ luminosity is $(3.2 \pm 0.3) \times10^{36}\;\text{erg}$ $\text{s}^{-1}$.  We can compare this luminosity to the X-ray luminosity to evaluate whether the AGN could be the source of the observed Br$\gamma$ line or not.  The best fit power law model for the source gives an X-ray luminosity is $L_{2-10} = 2.7 \times 10^{39}\;\text{erg}$ $\text{s}^{-1}$  \citep[][]{Reines2011}.  This best fit also gives $N_{H,\text{pow}}=6.3^{+5.5}_{-3.6} \times 10^{22} \text{cm}^{-2}$ which corresponds to $A_V \sim 14 - 60$.   We can convert the X-ray luminosity to an expected Br$\gamma$ luminosity; \citet[][]{Panessa2006A&A} derives an empirical linear relationship of X-ray to $H_{\alpha}$ luminosity for a sample of 47 AGNs and 30 QSOs: $\log_{10}L_X=(1.06\pm0.04)\log_{10}L_{\alpha}-(1.15\pm 1.85)$; it appears that this error bar is misprinted; we recalculated the error on the relation by examining the scatter in the data and found a zeropoint error of 0.85, therefore we assume a relation $\log_{10}L_X=(1.06\pm0.04)\log_{10}L_{\alpha}-(1.15\pm 0.85)$. We also note that this relation may not hold at low accretion rates or at low masses (e.g., for Henize~2-10).  Proceeding with this caveat, the predicted $L_{H\alpha}$ luminosity ranges from $7.56 \times 10^{38}\;\text{erg}$ $\text{s}^{-1}$ to $9.38\times 10^{38}\;\text{erg}$ $\text{s}^{-1}$. Next, assuming case B recombination and a temperature of $T=10,000$ K, the ratio of Br$\gamma$/H$_\alpha$ is 0.01  \citep[][]{Osterbrock1989}.  Thus the predicted Br$\gamma$ luminosity is between $7.56\times 10^{36}\;\text{erg}$ $\text{s}^{-1}$ and $9.38\times 10^{36}\;\text{erg}$ $\text{s}^{-1}$.  The error in this determination is dominated by the error on the conversion between $L_X$ and $L_{H\alpha}$; the conversion from H$\alpha$ to Br$\gamma$ is similar over the full range of reasonable temperatures (i.e.,~5,000--20,000~K).  

We now estimate whether this predicted Br$\gamma$ luminosity is consistent with the observed luminosity given the extinction values from the X-ray data and observed from line-emission ratios by \citet[][]{Cresci2010A&A}.  With the observed and the predicted Br$\gamma$ luminosity above, the ratio $L_{\text{Br}\gamma,\text{pre}}/L_{\text{Br}\gamma,\text{obs}}\sim 2.4 - 3.0$; therefore, assuming the \citet[][]{Cardelli1989ApJ} extinction law, the extinction suggested is $A_V \sim 8 - 12$ or corresponding to a hydrogen column of $N_{H}=1.5^{+0.75}_{-0.00}\times10^{22}\text{cm}^{-1}$ \citet[][]{Bohlin1978ApJ}. The extinction maps of \citet[][]{Cresci2010A&A} show values between $A_V \sim 2-14$; nearest the location of the AGN, the average extinction is approximately $A_V = 10.6$.  The hydrogen column also overlaps with the poorly constrained Hydrogen column fit from the X-ray spectrum by \citet[][]{Reines2011} of $N_{H,\text{pow}}=6.3^{+5.5}_{-3.6} \times 10^{22} \text{cm}^{-2}$.   Thus it is plausible that the Br$\gamma$ emission we observe is coming from accretion onto the BH.

If the Br$\gamma$ emission is indeed from the BH, it allows us to measure the velocity of the BH relative to the galaxy as a whole.  Fitting the spectrum shown in Fig.~\ref{AGN} (with r = $0.2\arcsec$), the Br$\gamma$ radial velocity is ($28.3 \pm 1.5)$~km~$\text{s}^{-1}$ redshifted relative to systemic velocity of the galaxy and its velocity dispersion is ($23.7 \pm 0.7$)~km~$\text{s}^{-1}$ taking into account the local instrumental dispersion.  We assume the systemic velocity of the galaxy is $\sim$872$\pm$6~km~$\text{s}^{-1}$ \citep[][]{Marquart2007A&A}. 
%%%%%%%%%%%%%%%%%%%%%%%%%%%%%%%%%%%%%%%%%%%%%%%%%%%%%%%%%%%%%%%%%%%%%%%%%%%%%
% CONCLUSION SECTION
%%%%%%%%%%%%%%%%%%%%%%%%%%%%%%%%%%%%%%%%%%%%%%%%%%%%%%%%%%%%%%%%%%%%%%%%%%%%%
\section{Conclusions}\label{sec:conclusion}
We have examined the overall morphology and nuclear regions of Henize~2-10 using ground-based photometric data, adaptive optics Gemini/NIFS data and {\it HST} data.  Our primary findings are: 
\begin{enumerate}

\item Henize~2-10 is well fit by a two component S\'ersic profile.  The inner S\'ersic profile has $n \sim 0.6$ and $r_{\text{eff}}\sim6\arcsec$ ($\sim 258$ pc), and the outer S\'ersic profile has $n \sim 1.8$ and $r_{\text{eff}}\sim25\arcsec$ ($\sim$~1kpc).  The absolute magnitude of Henize 2-10 within 4.3 kpc is $M_g = -19.2\pm0.2$, $M_r = -19.8\pm0.4$ in $g$-band and $r$-band.

\item The total stellar mass of Henize~2-10 within 4.3 kpc is $(10\pm3)\times10^{9}M_{\odot}$ as derived from the $r$-band, a factor of $\sim3$ times higher than previous estimates based on the $K$-band luminosity.

\item Apart from the inner starburst, Henize~2-10 appears to be typical early-type galaxy.  Its outer color $(g-r)_0 = 0.75$ is consistent with an old population.  Furthermore, it is non-rotating and dispersion-dominated near the center and has an outer S\'ersic index consistent with other early-type galaxies of similar luminosity.

\item We estimate the dynamical mass of the brightest SSC in Henize~2-10 to be $2.3\pm0.6 \times10^6$~M$_\odot$, higher than previous estimates.  We use this mass estimate to derive new masses for 11 clusters at the center of Henize~2-10, assuming a constant $M/L$ for all the clusters.

\item The timescale for dynamical friction is $<10^9$ years for the SSCs at the center of Henize~2-10.  An NSC of mass $>10^6$~M$_\odot$ should be formed within a few hundred Myr; this cluster mass would be typical for galaxies of Henize~2-10's mass.  Thus, we are seeing NSC formation in progress in this galaxy, and this formation appears to be independent of the formation of the BH in this system.

\item  While there are few examples of $M_{\star}<\sim 10^{10} M_{\odot}$ star-forming galaxies with observational evidence for a massive BH \citep[e.g.,][]{Greene2012NatCo, Schramm2013ApJ, Reines2014ApJ}, the occupation fraction of massive BHs in early-type galaxies of this mass is likely $\sim$90\% \citep[][]{Miller2014}.  Therefore, it is reasonable that Henize 2-10 hosts a massive BH.  Our observations allow us to place a firm upper limit on the BH mass of $M_{\bullet} < 10^7 M_{\odot}$.  We do not detect coronal emission from the BH, but Br gamma emission consistent with the X-ray emission is detected at the location of the BH. 

\end{enumerate}
      
\acknowledgments
We would like to thank Maureen Conroy at the SAO/TDC for providing us the Magellan/Megacam data of $g$-, $i$-, and $r$-bands, Micheva Genoveva for generously sharing us with her SB profiles in 5 Bessel-Johnson filters $UBVHK$, and the University of Utah, Physics and Astronomy Department for supporting this work.  The authors also thank Fabio Antonini for helpful discussions. Support for Amy E. Reines was provided by NASA through the Einstein Fellowship Program, grant PF1-120086.

%%%%%%%%%%%%%%%%%%%%%%%%%%%%%%%%%%%%%%%%%%%%%%%%%%%%%%%%%%%%%%%%%%%%%%%%%%%%%
% BIBLIOGRAPHY SECTION.
%%%%%%%%%%%%%%%%%%%%%%%%%%%%%%%%%%%%%%%%%%%%%%%%%%%%%%%%%%%%%%%%%%%%%%%%%%%%%
%\bibliographystyle{apj}
%\bibliography{He210}

\end{document}